\newcounter{myctr}

\documentclass{ws-acs}

\usepackage[compress]{cite}
\usepackage{xcolor}
\usepackage{subcaption}
\usepackage[verbose]{hyperref}
\hypersetup{colorlinks=false,allbordercolors=blue,pdfborderstyle={/S/U/W 1}}

\begin{document}

\makeatletter
\def\@biblabel#1{[#1]}
\makeatother

\markboth{Mauro, Pedreschi, Lambiotte, Pappalardo}{Dynamic models of Gentrification}

\catchline{}{}{}{}{}
%
%%%%%%%%%%%%%%%%%%%%%%%%%%%%%%%%%%%%%%%%%%%%%%%%%%%%%%%%%%%%%%%%%%%%

\title{Dynamic models of gentrification}

\author{Giovanni Mauro}
\address{ISTI-CNR, Pisa, Italy\\
Scuola Normale Superiore, Pisa, Italy\\
Department of Computer Science, University of Pisa, Pisa, Italy\\
IMT School for Advanced Studies, Lucca, Italy\\
\email{giovanni.mauro@sns.it}}

\author{Nicola Pedreschi}
\address{Mathematical Institute, University of Oxford, United Kingdom\\
\email{nicola.pedreschi@maths.ox.ac.uk}}

\author{Renaud Lambiotte}
\address{Mathematical Institute, University of Oxford, United Kingdom\\
\email{renaud.lambiotte@maths.ox.ac.uk}}

\author{Luca Pappalardo}
\address{ISTI-CNR, Pisa, Italy\\
Scuola Normale Superiore, Pisa, Italy\\
\email{luca.pappalardo@isti.cnr.it}}

\maketitle

\begin{abstract}
The phenomenon of gentrification of an urban area is characterized by the displacement of lower-income residents due to rising living costs and an influx of wealthier individuals. This study presents an agent-based model that simulates urban gentrification through the relocation of three income groups -- low, middle, and high -- driven by living costs. The model incorporates economic and sociological theories to generate realistic neighborhood transition patterns. We introduce a temporal network-based measure to track the outflow of low-income residents and the inflow of middle- and high-income residents over time. Our experiments reveal that high-income residents trigger gentrification and that our network-based measure consistently detects gentrification patterns earlier than traditional count-based methods, potentially serving as an early detection tool in real-world scenarios. Moreover, the analysis highlights how city density promotes gentrification. This framework offers valuable insights for understanding gentrification dynamics and informing urban planning and policy decisions.
\end{abstract}

\section{Introduction}
\label{sec:intro}

Cities are dynamic systems\cite{batty2013new,jacobs1961death,pappalardo2023future} in which the interactions between numerous agents determine the emergence of non-trivial patterns at different scales, such as traffic congestion\cite{cornacchia2022routing}, epidemic spreading\cite{lucchini2021living,oliver2020mobile}, and socioeconomic segregation\cite{schelling1971dynamic,gambetta2023mobility,moro2021mobility}.
Gentrification, first defined by Ruth Glass in 1964\cite{glass2013aspects}, describes the transformation of neighborhoods from working-class to affluent areas, often displacing original residents, and potentially undermining urban diversity and affordability\cite{jacobs1961death}. Slater\cite{slater2011gentrification} divides gentrification research into two main strands: production-side and consumption-side theories. Both reject the view that gentrification is a benign return to urban centers\cite{lipton1977evidence, leroy1983paradise}, with production-side theories linking it to economic factors such as the rent gap\cite{smith1979toward, smith1987gentrification}and housing quality decline\cite{lowry1960filtering}. Conversely, consumption-side theories emphasise the growing appeal of city centers\cite{simpson2017planning}, where proximity to urban amenities fuels demand. Ley\cite{ley2003artists} argues that artists, drawn by the cultural and social vitality of these areas, act as early catalysts for gentrification, eventually attracting wealthier residents and driving up property values\cite{hamnett2000gentrification}.

%Economic%
Computational studies have blurred the lines between these perspectives, focusing on housing market dynamics, particularly fluctuations in rent and housing prices, as these reflect the real-world data used to validate their models. O'Sullivan\cite{o2002toward} proposes a model incorporating housing markets, social networks, history, and policies, illustrating how gentrification operates in cycles influenced by these factors. Redfern\cite{redfern1997new} introduces the ``investment gap," emphasizing the difference between non-modernized homes and their potential if modernized, with domestic technologies driving gentrification. Other computational models simulate household movements\cite{liu2016abstract}, considering vacancies, accessibility, socioeconomic status, and urban policies\cite{torrens2007modeling, eckerd2019gentrification}. Alternatively, machine learning approaches have been implemented in an attempt to predict gentrification events\cite{rigolon2019toward,reades2019understanding}, in some limited cases taking into account proxies for human mobility in urban areas\cite{gardiner2021mobility}. A recent work by Shaw et al.\cite{shaw2024dynamical} showed how even a simple dynamical system model of gentrification, focused on neighborhood attractiveness and artist populations, can generate complex temporal patterns including synchronized oscillations and transient chaos.

In this study, we present an agent-based model of gentrification, inspired by the work of Schelling on urban segregation\cite{schelling1971dynamic}. Rather than focusing on replicating housing market dynamics, we base our analysis of gentrification on the relocation flows of citizens in a stylised urban grid. Our model is founded on the key assumption that gentrification is driven by socioeconomic inequality and additionally by differing relocation strategies across income levels. 
As Nieuwenhuis et al.\cite{nieuwenhuis2020does} found, in high-inequality contexts, low-income groups are confined to deprived areas while high-income groups maintain spatial dominance, creating a ``vicious circle" that perpetuates inequality.
Based on further insights from the literature\cite{christafore2019neighbourhood, banabak2024gentrification, hu2024corporate}, we assume that agents relocate according to their socioeconomic conditions\cite{mantzaris2020incorporating}: low-income agents move when priced out of a neighbourhood, medium-income agents gravitate toward areas with similar economic conditions and quality of life, while high-income agents are attracted to areas undergoing economic growth where they can maximise investment returns. 
These contrasting behaviours, along with a heavy-tailed income distribution, are the only drivers of neighbourhood transformations. While our agent-based model builds on simple rules, it generates complex and emergent dynamics, requiring a rigorous complex systems approach to quantify and interpret the multifaceted aspects of gentrification.

Within our modeling framework, we show that gentrification emerges only when high-income residents have some mobility, even if minimal, highlighting how their movement patterns catalyse the process. We treat relocation flows of agents in our city as time-varying edges in a temporal network\cite{masuda2016guide,gueuning2020rock}, leveraging established tools from network science and human mobility research\cite{barbosa2018human,noulas2012tale,pappalardo2024survey,mauro2022generating, pappalardo2023future}.
We introduce two novel measures to quantify gentrification within our theoretical framework. The first measure translates the conventional definition of gentrification into a metric based on the over-representation of middle- and high-income agents in a given area. The second measure captures the dynamics of gentrification by tracking the inflow of these agents alongside the simultaneous outflow of low-income residents. Our findings demonstrate that this dynamic measure can consistently detect gentrification earlier than the more intuitive, count-based approaches, making it a potential early-warning indicator for policymakers aiming to mitigate its impacts. Additionally, our framework can simulate the effects of various urban planning strategies and city characteristics on gentrification. Notably, we observe a direct correlation between urban density and the frequency of gentrification events. Overall, our model and measures offer a comprehensive perspective on gentrification, shedding new light on this complex urban phenomenon.

\section{Agent-based model of gentrification}
\label{sec:model}

\subsection{ Modelling the city and its citizens.} 

\begin{figure}[htb!]
\centering
    \includegraphics[width=0.8\textwidth]{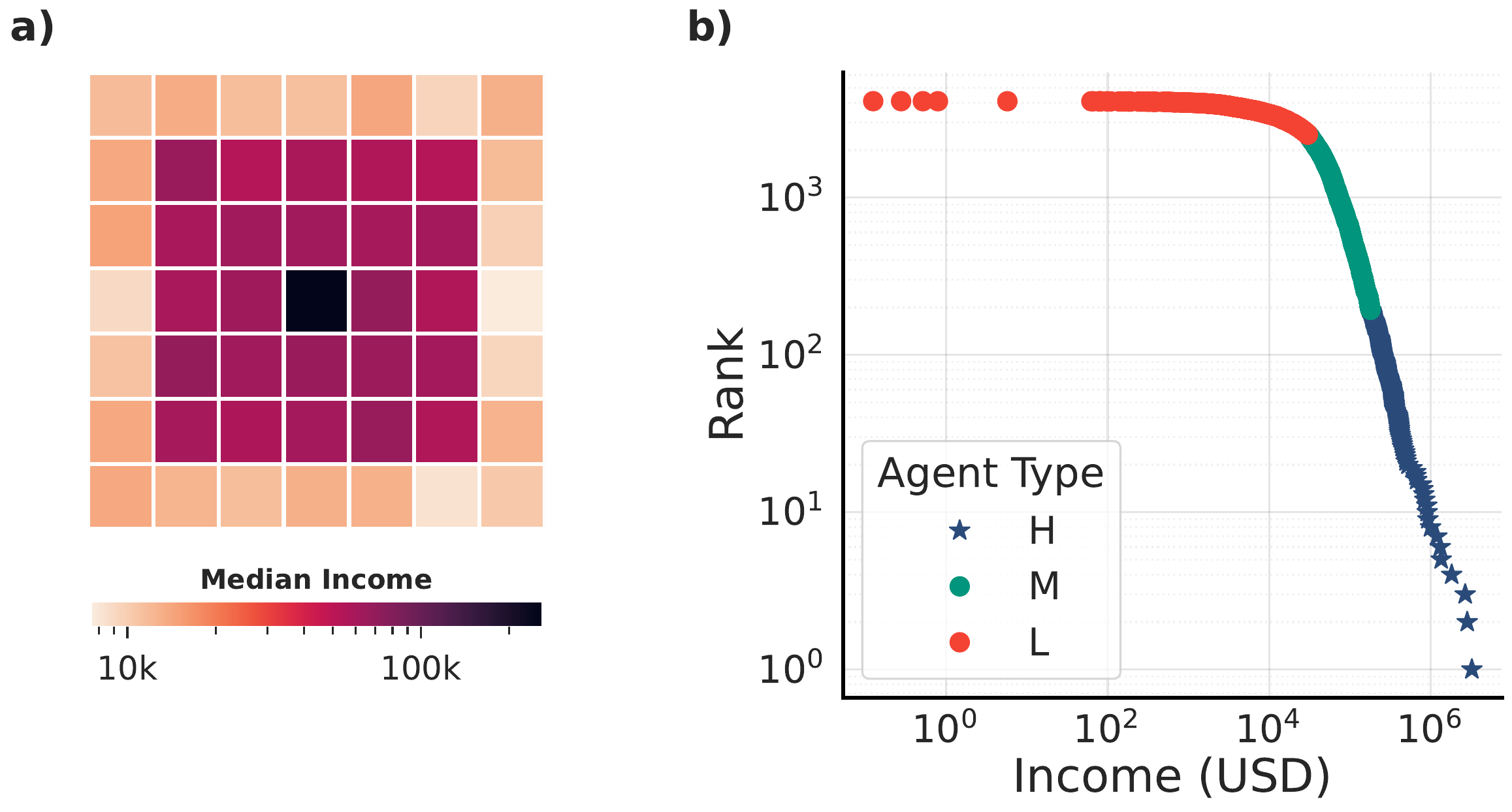}
    \caption{\textbf{Income and spatial distribution:}  \textbf{a)} Heatmap reporting the median income of each cell at the beginning of a simulation. The darker the colour, the higher the median income of the cell. \textbf{b)} Log-log plot of the Initial distribution of the income, sampled from the 2022 USA Social Security Administration report\cite{ssa2022compensation}, of the agents in a simulation with N = $2^{12}$ agents. } 
    \label{fig:income_heat}

\end{figure}

We model the urban environment as a 7×7 grid, with each cell representing a city neighbourhood (Figure \ref{fig:income_heat}a). We populate this grid with $2^{12}$ agents, categorised into three socioeconomic groups: low-income ($L$), middle-income ($M$), and high-income ($H$).
At the beginning of a simulation, each agent is assigned a fixed income $w$, sampled from real-world data, and all agents are  divided into three groups according to their assigned incomes: $L$ accounting for 38\% of the population; $M$ accounting for 57\% of the population; and  $H$  corresponding to the remaining 5\% of the population of the model city.
The method of income assignment is based on data from the 2022 USA Social Security Administration report\cite{ssa2022compensation} (see Methods for further details) and allows income variation within each agent class.
Figure \ref{fig:income_heat}b shows one realisation of income assignment to the agents resulting in a heavy-tailed agent income distribution (more details in Methods).

The spatial distribution of the agents follows a socioeconomic radial gradient: $H$ agents predominantly occupy the city centre, $M$ agents populate the inner areas, and $L$ agents are concentrated in the periphery (Figure \ref{fig:income_heat}a). This mono-centric structure reflects the presence of a dominant central business district, found in many cities or metropolitan areas of varying sizes\cite{barthelemy2016structure,arribas2014validity}. 
Each cell (neighbourhood) $j$ has a fixed maximum capacity $K$, which limits its occupancy, that is, how many agents can stay there. When $n(j)$, the number of agents in the cell $j$, reaches maximum capacity $K$, agents are allocated to the nearest available locations. 
Our model operates based on two key parameters: the parameter $p_H$ that corresponds to the probability that a $H$ agent relocates from its current cell to a new one where the median agent-income is increasing;  the parameter $\theta$ that is the width of the time window over which $H$ agents evaluate cell growth trends. Our model simulates a 7x7 grid city comprising 49 neighbourhoods, a scale consistent with moderately large urban areas. To ensure the robustness of our findings, we extended our analysis to an even larger urban environment (9$\times$9 grid, 81 neighbourhoods), as well as to a city with maximally mixed distribution of agents of the three types, where the radial gradient of income is lost, obtaining similar results (see Supplementary Information 3-5).

\begin{figure}[htb!]
\centering
    \includegraphics[width=0.75\textwidth]{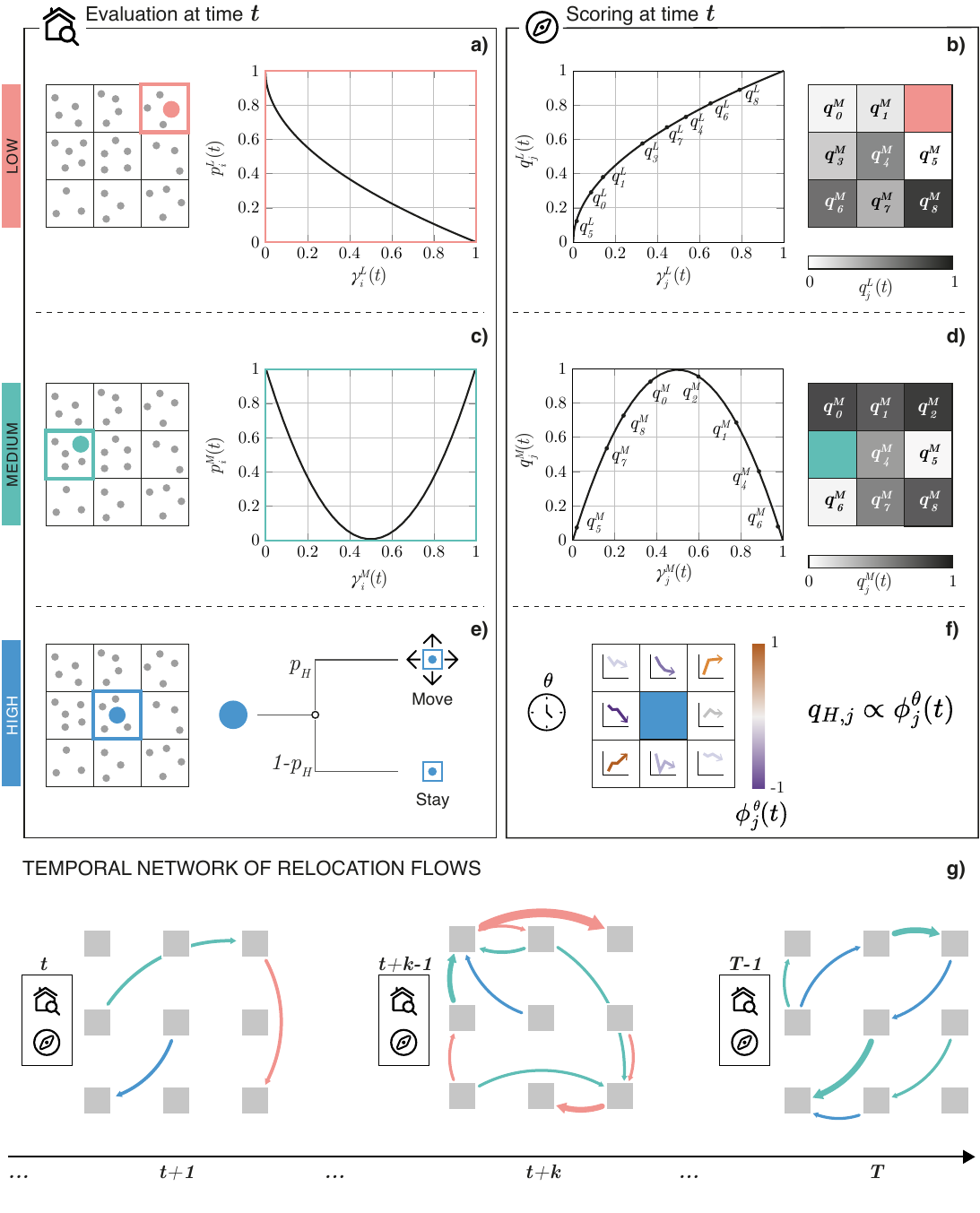}
    \caption{\textbf{Income-profile-specific agent behaviour}:\small{
The city is represented as a grid at time $t$, with three types of agents distributed across the cells. We track a low-income agent $L$ (red cell), a middle-income agent $M$ (green cell), and a high-income agent $H$ (blue cell). \textbf{(a)} The $L$ agent moves from its current cell $i$ with a probability $p^L_i(t)$, which is inversely proportional to its income percentile in $i$. \textbf{b)} When $L$ moves, it relocates to a new cell $j$ with probability $q^L_j(t)$. \textbf{(c)} The $M$ agent moves from its current cell $i$ with probability $p^M_i(t)$, that its higher the more the income percentile of $M$ is extreme. \textbf{d)} When $M$ moves, it relocates to a new cell $j$ with probability $q^M_j(t)$. \textbf{(e)}  The $H$ agent moves with a fixed probability $p_H$. \textbf{f)} The $H$ agent moves to a new cell $j$ based on its growth rate $\phi^\theta_j(t)$. At time $t+1$, the $L$ agent moves to cell $j=8$, the $M$ agent to cell $j=2$, and the $H$ agent to cell $j=6$. \textbf{g)} The output of our model is a temporal network, where nodes are grid cells and edges represent agent flows between cells over time. The first snapshot of the network at $t+1$ shows the flows generated by \textbf{a-f)}. The model stops when no $L$ agent can move, resulting in no red edges in the final network snapshot at time $T$.}}
\label{fig:fig_0}
\end{figure}
\vspace{4cm}
\subsection{Agent relocation dynamics.} 
Our model implements income-dependent relocation rules for three agent classes across a set of cells, where each cell $i$ has a maximum capacity $K$. Let $n(i)$ denote the number of agents in cell $i$ at any given time. Figure \ref{fig:fig_0}a-f illustrates the rules behind agents' behaviours.

A $L$ agent is likely to move away from a cell if they are significantly poorer than other agents in the cell. The agent moves, with higher probability, to a cell where the income gap with current residents is smaller.  The probability of a $L$ agent leaving cell $i$ at time $t$, shown in Figure \ref{fig:fig_0}a, is given by:
\begin{equation}
    p^L_i(t) = 1 - \gamma^L_i(t)^{\frac{1}{2}},
\end{equation}
where $\gamma^L_i(t) \in (0,1]$ represents the agent's relative income percentile within cell $i$'s income distribution at time $t$. Upon deciding to move, $L$ agents select from the set $\mathcal{J}$ of available cells, defined as cells where $n(j) < K$.  As illustrated in Figure \ref{fig:fig_0}b, the probability of selecting cell $j$ is:
\begin{equation}
    \rho^L_j(t) = \frac{q^L_j(t)}{\sum_{n \in \mathcal{J}}q^L_n(t)}, \quad q^L_j(t) = \gamma^L_j(t)^{\frac{1}{2}},
\end{equation}
where $q^L_j(t)$ is the cell's attractiveness score based on the agent's prospective income percentile in cell $j$.

A $M$ agent is likely to move away when its income significantly deviates from the median income in the current cell. 
The agent will move to a cell where the gap with the median income is lower. Their relocation probability, visualized in Figure \ref{fig:fig_0}c, follows::
\begin{equation}
    p^M_i(t) = 4(\gamma^M_i(t)-0.5)^2,
\end{equation}
where $\gamma^M_i(t) \in [0,1]$ is the $M$ agent's relative income percentile in cell $i$.  Figure \ref{fig:fig_0}d shows how their destination selection probability is determined by:
\begin{equation}
    \rho^M_j(t) = \frac{q^M_j(t)}{\sum_{n \in \mathcal{J}}q^M_n(t)}, \quad q^M_j(t) = 1 - 4(\gamma^M_j(t)-0.5)^2,
\end{equation}
where $q^M_j(t)$ scores cells based on proximity to their median income.

For both $L$ and $M$ agents, we choose nonlinear functional forms (square root for $L$, quadratic for $M$) to ensure that small differences in percentiles lead to larger differences in probabilities when agents are far from their preferred positions. The denominators in $\rho^L_j(t)$ and $\rho^M_j(t)$ serve as normalisation factors, ensuring proper probability distributions.

A $H$ agent moves from its current cell $i$ with a fixed probability $p_H \in [0,1]$. Let $\Tilde{w}_{j}(t)$ denote the median income in cell $j$ at time $t$. $H$ agents select from set $\mathcal{H}$ of cells satisfying both $n(j) < K$ and $\phi_j^\theta(t) > \phi_i^\theta(t)$, where $\phi_j^\theta(t)$ is the average growth rate of median income over the past $\theta$ time steps, as reported in Figure \ref{fig:fig_0}e-f:
\begin{equation}
    \phi_j^\theta(t) = \frac{1}{\theta}\sum_{\tau=t-\theta}^{t}[\Tilde{w}_{j}(\tau)-\Tilde{w}_{j}(\tau-1)].
\end{equation}
Their destination probability is:
\begin{equation}
    q^H_j(t) = \frac{\phi_j^\theta(t)}{\sum_{n\in \mathcal{H}}\phi_n^\theta(t)}.
\end{equation}
The definition of the destination probability $q^H_j(t)$ reflects the profit-orientated nature of the movement of H agents. Cells that underwent a higher growth in the last $\Theta$ steps are thus more likely to be selected as destination.
To ensure an easier interpretation of the results of our model, we impose that H agents do not operate (move) in the first $\theta$ steps, i.e., until a full evaluation of the growth-rate $\phi^\theta(t)$ of cells has been completed.
The simulation terminates at time $T$ when all $L$ agents can only find cells that would place them in the lowest income percentile, or after 300 time steps. We denote the termination time as T. All simulations were implemented using the Python library \texttt{mesa}\cite{python-mesa-2020}. The complete source code for model implementation and experimental procedures is available at:
\url{https://github.com/mauruscz/Gentrification}

\section{Measures of gentrification}
\label{sec:measures}

Our gentrification model generates dynamic flows of agents of the three types among grid cells (see Figure \ref{fig:fig_0}g). We model these time-varying flows as a dynamic network where nodes represent grid cells and edges correspond to movements of agents of the three types. The network consists of three layers, each representing flows of $L$, $M$, or $H$ agents. 
This representation results in a temporal network\cite{Holme:2012,holme2015modern,masuda2016guide},which is multi-layer (one per each edge type)\cite{kivela_multilayer} and  weighted\cite{Barrat_weighted_complex}, defined as follows:
\begin{equation}
    \text{G}^\alpha(t) = \Big\{ A_{i,j}^\alpha(t) | i,j \in V; \alpha \in \{L, M, H\}; t \in [0,T]\Big\}
\end{equation}
\noindent where $V$ denotes the set of nodes (grid cells) in the network, $\alpha$ represents the layer index corresponding to $L$, $M$ or $H$ agents, and $A^\alpha(t)$ is the adjacency matrix corresponding to the network in layer $\alpha$ at time $t$. 
The matrix elements of $\mathbf{A}^\alpha(t)$ correspond to weighted, directed edges connecting node pairs in layer $\alpha$, representing the relocation flows of agents of type $\alpha$ between contiguous time steps of the agent-based model.
For each layer $\alpha \in [L,M,H]$ at time $t$, we define three quantities: $n^i_{\alpha}(t)$, the number of agents in node $i$, and $s^{\text{in}}_{i,\alpha}(t)$ and $s^{\text{out}}_{i,\alpha}(t)$, the in- and out-strength of node $i$, respectively. The latter two quantify the flow of agents in $[L,M,H]$ moving to or from node $i$ between $t-1$ and $t$:
\begin{equation}
\begin{split}
s^{\text{in}}_{i,\alpha}(t) &= \sum_i A^{\alpha}_{i,j}(t), \\
s^{\text{out}}_{i,\alpha}(t) &= \sum_j A^{\alpha}_{i,j}(t), \\
\alpha &= L, M, H.
\end{split}
\label{eq:strengths}
\end{equation}
where $A^{\alpha}_{i,j}(t)$ represents the $(i,j)$ element of the weighted adjacency matrix for layer $\alpha$ at time $t$.

In the following subsections, we introduce two complementary metrics to capture the dynamics of gentrification. The first metric adopts a ``migration stock" perspective \cite{abel2019bilateral}, aiming to formalise the typical definition found in the social science literature, which largely emphasises changes in neighbourhood composition over time. These definitions \cite{glass2013aspects, ley2003artists, smith1987gentrification} generally focus on the transformation of a place, such as the increasing presence of higher-income residents, rather than on the movement of people itself. To complement this, the second metric takes a ``migration flow" perspective \cite{abel2019bilateral}, shifting attention to the dynamics of displacement. It specifically measures the number of $L$ agents who leave a place as a result of being pushed out by incoming $M$ and $H$ agents.

\subsection{Count-based measure.} 
Gentrification is often defined as a period in which a neighbourhood (a cell in the grid  and a node in the temporal network), previously populated by a majority of lower-income citizens, undergoes a gradual replacement of its population with middle- and higher-income citizens\cite{glass2013aspects,jacobs1961death}. 
To capture this process, we introduce a measure of gentrification for each node in the network, denoted as $\mathcal{G}_{count}^i(t,\Delta)$, based on the number of agents from each socioeconomic class present in node $i$ at two time points: the current time $t$ and an earlier time $t-\Delta$. For each time point, we consider the counts of $H$, $M$ and $L$ agents. At time $t$, these counts are represented by $n^i_H(t)$, $n^i_M(t)$, and $n^i_L(t)$, respectively. Similarly, at time $t-\Delta$, the counts are denoted by $n^i_H(t-\Delta)$, $n^i_M(t-\Delta)$, and $n^i_L(t-\Delta)$. The definition of $\mathcal{G}_{count}^i(t,\Delta)$ is as follows:
\begin{equation}
\mathcal{G}_{count}^i(t,\Delta) = \frac{1}{\Delta}\sum_{\tau=t-\Delta}^{t}\frac{n^i_H(\tau)+n^i_M(\tau)}{n^i_H(\tau)+n^i_M(\tau)+n^i_L(\tau)}
\end{equation}
$\mathcal{G}_{count}^i(t,\Delta)$ represents the average fraction of $H$ and $M$ agents in node $i$ over the time window $[t-\Delta,t]$. We establish a significance threshold $n_{H,M}^*$, defined as the expected node-wise fraction of $H$ and $M$ agents under uniform random distribution across the grid.
Values of $\mathcal{G}_{count}^i(t,\Delta)$ exceeding $n_{H,M}^*$ indicate an over-representation of $H$ agents. Gentrification is identified when $\mathcal{G}_{count}^i(t,\Delta)$ transitions from below to above this threshold. This metric functions as a node property, independent of the network edge dynamics.

To identify gentrification, we establish a critical threshold $n_{M,H}^*=0.62$, representing the city-wide proportion of $M$ and $H$ agents.
To precisely capture gentrification events, we introduce the binary indicator $\mathcal{G}_{bin}$:
\begin{equation}
    \mathcal{G}_{bin}(t,\Delta)=\begin{cases}
    1, & \text{if $\mathcal{G}_{count}(t,\Delta)>n_{H,M}^*$}.\\
    0, & \text{otherwise}.
  \end{cases}
\end{equation}
We define $t_{shift}$, the onset of a gentrification event, as the moment when $\mathcal{G}_{bin}$ shifts from 0 to 1, indicating that a neighborhood has crossed the critical $M,H$ population threshold. More mathematical details about $t_{shift}$ can be found in Methods.

\subsection{Network-based measure.}

The count-based measure, while providing an intuitive quantification of gentrification, has notable limitations: it requires a significance threshold and only detects completed transitions, disregarding inter-neighbourhood dynamics. To overcome these constraints, we define a gentrification measure based on the temporal relocation network, $\mathcal{G}_{net}^i(t,\Delta)$ that captures relocation patterns between neighborhoods. For any node $i$ in the network, this measure considers the net-outflow of $L$ agents, $\varphi^{\text{out}}_i(t)$, and the net-inflow of $M$ and $H$ agents, $\varphi^{\text{in}}_i(t)$:
\begin{equation}
\varphi^{\text{out}}_i(t) \equiv  \frac{s^{\text{out}}_{i,L}(t)-s^{\text{in}}_{i,L}(t)}{s^{\text{out}}_{i,L}(t)+s^{\text{out}}_{i,M}(t)+s^{\text{out}}_{i,H}(t)}
\label{eq:flowout}
\end{equation}
\begin{equation}  \varphi^{\text{in}}_i(t) \equiv \frac{s^{\text{in}}_{i,M,H}(t)- s^{\text{out}}_{i,M,H}(t)}{s^{\text{in}}_{i,L}+s^{\text{in}}_{i,M}(t) + s^{\text{in}}_{i,H}(t)}
    \label{eq:flowin}
\end{equation}

\noindent where $ s^{\text{in}}_{i,M,H}(t) = s^{\text{in}}_{i,M}(t) + s^{\text{in}}_{i,H}(t)$ and $s^{\text{out}}_{i,M,H}(t) = s^{\text{out}}_{i,M}(t) - s^{\text{out}}_{i,H}(t)$.

$\varphi^{\text{out}}_i(\tau)$ is high when the outflow of $L$ agents from node $i$ corresponds to a high fraction of the overall outflow of agents from $i$, i.e., the denominator in Equation (\ref{eq:flowout}); $\varphi^{\text{in}}_i(\tau)$ is high when the inflow of $M$ and $H$ agents to node $i$ corresponds to a high fraction of the overall inflow of agents towards $i$, i.e., the denominator in Equation (\ref{eq:flowin}). $\mathcal{G}_{net}^i(t,\Delta)$ is therefore defined as the geometric mean of the averages of $\varphi^{\text{out}}_i(t)$ and $\varphi^{\text{in}}_i(t)$ over the last steps $\Delta$:
\begin{equation}
\mathcal{G}_{net}^i(t,\Delta)\equiv \sqrt{\Bigg(\frac{1}{\Delta} \sum_{\tau=t-\Delta}^{t} \varphi^{\text{out}}_i(\tau)\Bigg) \cdot \Bigg(\frac{1}{\Delta} \sum_{\tau=t-\Delta}^{t} \varphi^{\text{in}}_i(\tau)\Bigg)}.
    \label{eq:Gnet}
\end{equation}
%The product of the averages over the period $[t-\Delta,t]$ is not equal to the average of the product $\varphi^{\text{out}}_i(t) \cdot \varphi^{\text{in}}_i(t)$ due to the different denominators in the flow definitions. 
In Equation (\ref{eq:Gnet}), we consider only positive values of $\varphi^{\text{out}}_i(t)$ and $\varphi^{\text{in}}_i(t)$. This approach ensures that high values of $\mathcal{G}_{net}^i(t,\Delta)$ occur only when both the outflow of $L$ agents and the inflow of $M$ and $H$ agents are substantial during the same period. Including negative values would erroneously indicate gentrification in areas where $L$ agents replace $M$ and $H$ agents, which actually signals neighbourhood impoverishment.

We define $t_{peak}$ as the time of a gentrification event, corresponding to a local maximum in $\mathcal{G}_{net}^i(t,\Delta)$ or the onset of a plateau after rapid growth (see Methods for details).

To quantify gentrification at the city scale, we introduce two cumulative metrics: $\mathcal{G}_{bin}^{city}$, the percentage of nodes experiencing transitions in $\mathcal{G}_{bin}$, and $\mathcal{G}_{net}^{city}$, the percentage of nodes showing peaks in $\mathcal{G}_{net}$ (see Methods for details).

\section{Results}
\label{sec:results}

\subsection{High income agents drive gentrification.}

We examine the impact of $H$ agent mobility on gentrification dynamics. Simulations were conducted with varying $H$ agent movement probabilities ($p_H$), while maintaining a fixed evaluation window of $\theta=20$ time steps for node growth rates.

\begin{figure}[]    
\centering
    \includegraphics[width=0.9\linewidth]{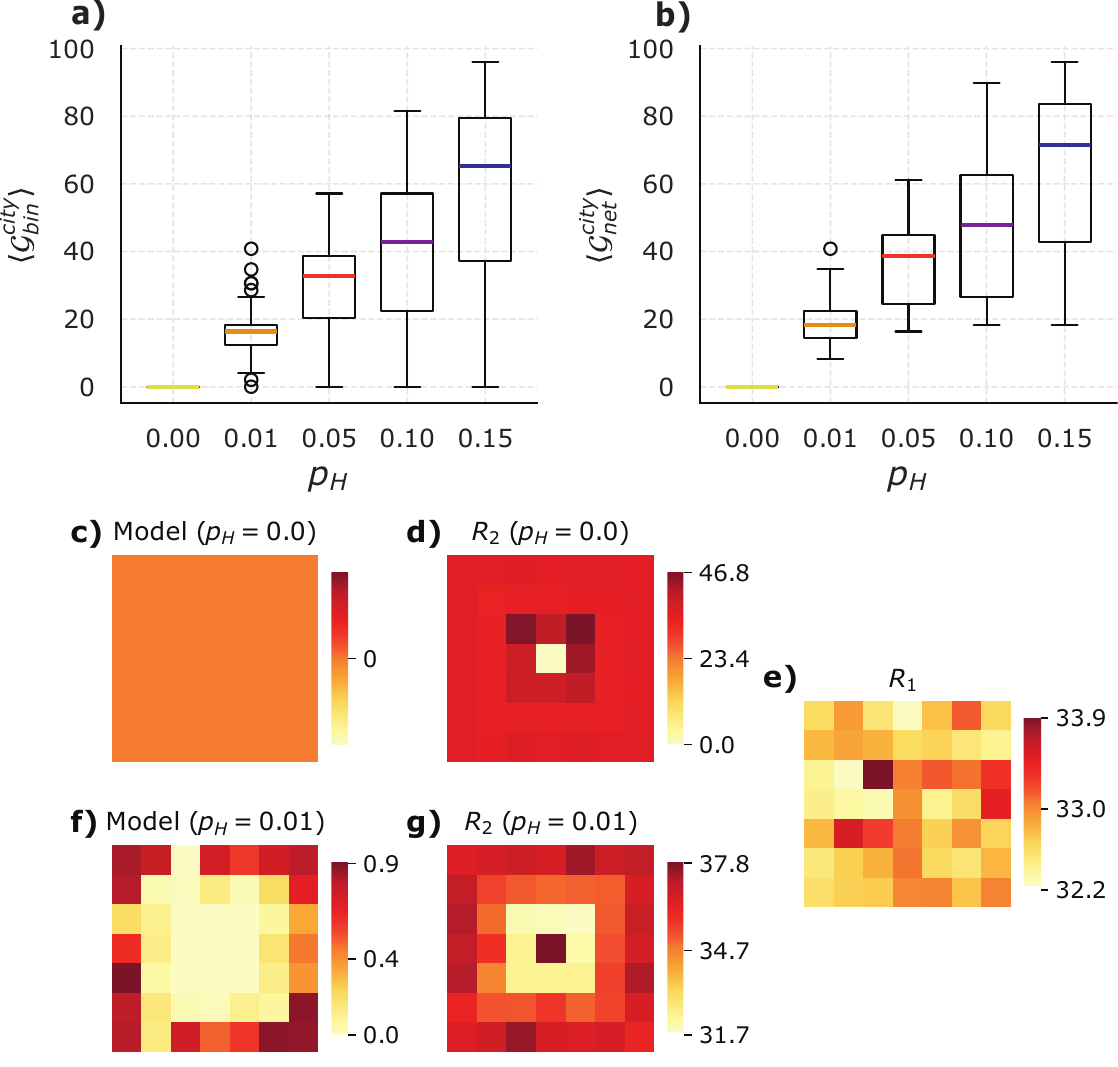}
    \caption{\textbf{Gentrification patterns in our model:}
    \textbf{a,b)} Gentrification level across 150 simulation of the model with a fixed $\theta=20$, $N=2^{12}$ and varying value of $p_H$ according to $\mathcal{G}_{bin}^{city}$ (\textbf{a}) and to the network based measure $\mathcal{G}_{net}^{city}$ (\textbf{b}). (\textbf{c-g}) Heatmaps of the average number of peaks of $\mathcal{G}_{net}^i(t)$ observed in any neighborhood $i$ over 150 simulations of our model (left), the $R_2$ null model (center) and the $R_1$ null model (right).
    }
    \label{fig:box_all}
\end{figure}

Figure \ref{fig:box_all}a,b illustrates the influence of $p_H$ on spatial gentrification patterns. With $p_H=0$, both $\mathcal{G}_{bin}^{city}$ and $\mathcal{G}_{net}^{city}$ yield 0\%, indicating complete absence of gentrification when $H$ agents are static. Introducing minimal $H$ agent mobility ($p_H=0.01$) triggers gentrification in 20-40\% of nodes, according to both metrics. This abrupt transition highlights the critical role of $H$ agent mobility in initiating the gentrification process. As $p_H$ increases, we observe a monotonic rise in gentrification levels, with $\mathcal{G}_{bin}^{city}$ and $\mathcal{G}_{net}^{city}$ showing similar trends but slightly different magnitudes.

To evaluate the significance of movement rules in driving gentrification, we compare our gentrification model with two null models: \emph{(i)} $R_1$, where agents decide whether and where to relocate with a fixed probability (50\%); \emph{(ii)} $R_2$, where only the destination choice is random (details in Supplementary Information 1). 
Figure \ref{fig:box_all}c-g shows heatmaps of the average $\mathcal{G}_{\text{net}}^i(t)$ peaks per neighbourhood for the gentrification model and the two null models.
Figure \ref{fig:box_all}c highlights that no gentrification occurs in our model when $p_H = 0.0$, while gentrification events persist in the null models (Figures \ref{fig:box_all}d,e) even in the absence of $H$ agents' movement. 
Figures \ref{fig:box_all}f shows how, in our model, with $p_H=0.01$, gentrification events are concentrated into peripheral neighbourhoods dominated by $L$ agents. In contrast, both null models  (Figures \ref{fig:box_all}e,g) show widespread gentrification events, including cyclical transitions between $L$- and $M$-majority in central neighbourhoods. Supplementary Figure S1 further shows that none of the null models converge to stable configurations.
The results indicate that the rules established by our model enhance the likelihood of gentrification occurring in neighbourhoods that are initially low-income. This pattern is not observed in the null models, highlighting the significance of the rules we developed for generating gentrification events, in line with the definition of gentrification found in the literature \cite{glass2013aspects,smith1987gentrification,simpson2017planning}.

\subsection{Network-based measure anticipates count-based measure.}
\begin{figure}[]
  \centering
  \includegraphics[width=.9\textwidth]{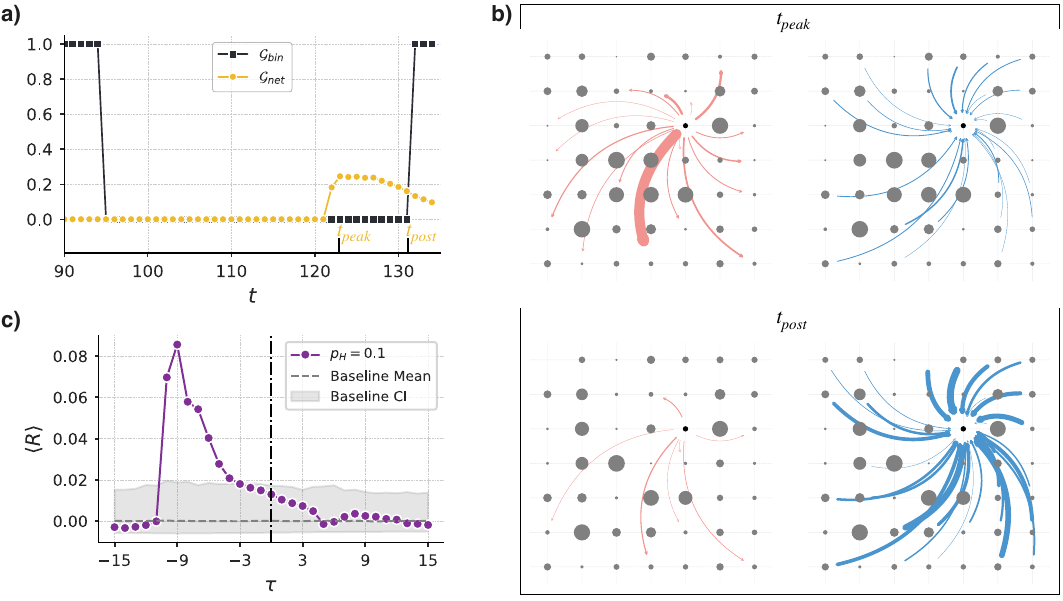}
  \caption{\textbf{Evolution of a gentrification event for a representative node ($p_H = 0.1$, $\theta = 20$, $\Delta=15$) and $N=2^{12}$}. \textbf{a)}: Comparison of $\mathcal{G}_{net}$ (in gold) and $\mathcal{G}_{bin}$ (in black) measures over time highlighting two key moments :  during and after the peak in $\mathcal{G}_{net}$, i.e.,  $t_{peak}$ and $t_{post}$, respectively. \textbf{b)}: Relocation in- and out-flows during (top), and after (bottom) the peak of $\mathcal{G}_{net}$. Red arrows correspond to $L$ agent outflows, blue arrows correspond to $M$ and $H$ agent inflows within $\Delta$, node size is proportional to agents population. (top) Peak: Simultaneous $L$ outflow and $M+H$ inflow. (bottom) Post-peak: Diminishing $L$ outflow, intensifying $M$ and $H$ inflow. Flows represent gross out- and in- strengths. \textbf{c):} Average cross-correlation $\langle R\rangle$ between $\mathcal{G}_{net}$ and $\mathcal{G}_{bin}$ across all nodes and 150 model runs, as a function of lag $\tau$. Grey area shows 95\% confidence interval; dotted line indicates null distribution mean.}
  \label{fig:preduringpost}
\end{figure}

While $\mathcal{G}_{bin}$ provides an intuitive measure of gentrification by capturing demographic transitions from $L$ to $M/H$ agent majorities, $\mathcal{G}_{net}$ enables earlier detection by identifying patterns of coordinated movement through temporal networks, revealing gentrification dynamics before visible demographic shifts occur.
Figure \ref{fig:preduringpost}a displays $\mathcal{G}_{net}(t,\Delta)$ and $\mathcal{G}_{bin}(t,\Delta)$ curves for a node in a representative simulation. In this example, the node initially experiences impoverishment, as $\mathcal{G}_{bin}$ transitions from 1 to 0 at approximately $t = 95$, indicating a shift from over- to under-representation of $M$ and $H$ residents. However, $\mathcal{G}_{net}$ shows a rapid increase at $t = 122$, followed by a plateau at $t = 123$, approximately 10 steps before the abrupt transition from 0 to 1 in $\mathcal{G}_{bin}$ at $t = 132$.

Figure \ref{fig:preduringpost}b illustrates relocation dynamics during and after the $\mathcal{G}_{net}$ peak for the same node (neighborhood) analyzed in Figure \ref{fig:preduringpost}a. Red arrows indicate $L$ agent outflows, and blue arrows show $M+H$ inflows. At the peak ($t_{peak}$), both occur simultaneously, with more $L$ agents leaving. After the peak, $M+H$ inflows increase while $L$ outflows decrease. Unlike the gross flows in the visualization, $\mathcal{G}_{net}$ captures net flows, offering a more nuanced understanding of these changes.

To verify the consistency with which $\mathcal{G}_{net}$ peaks precede $\mathcal{G}_{bin}$ transitions across our simulations, we conduct a lagged cross-correlation analysis between $\mathcal{G}_{net}$ and $\mathcal{G}_{bin}$ time series. Specifically, we first compute the cross-correlation for each node in the city grid, then average these cross-correlations across all nodes. This process is repeated for 150 independent model runs, and the results are again averaged to obtain the final cross-correlation profile, reported in Figure \ref{fig:preduringpost}c. We compute cross-correlations for lags $\tau \in [-15, +15]$ (see Methods for details), with significance tested against a null distribution. The highest correlation at $\tau = -9$ shows that $\mathcal{G}_{net}$ peaks systematically anticipate $\mathcal{G}_{bin}$ transitions by approximately 9 time steps. We further validate these findings repeating the analyses in the two randomized versions of the model. The anticipatory behaviour of $\mathcal{G}_{net}$ relative to $\mathcal{G}_{bin}$ transitions disappears in both the fully random and random-destination versions of the model  (see Supplementary Figure S2). This result support our main findings, demonstrating that $\mathcal{G}_{net}$ reliably anticipates $\mathcal{G}_{bin}$ only in scenarios where agents' movements are influenced by their income distribution of in the city neighbourhoods. More measures and parameter combination are provided in Supplementary Information 2.

\begin{figure}[htb!]
    \centering
    \includegraphics[width=0.65\linewidth]{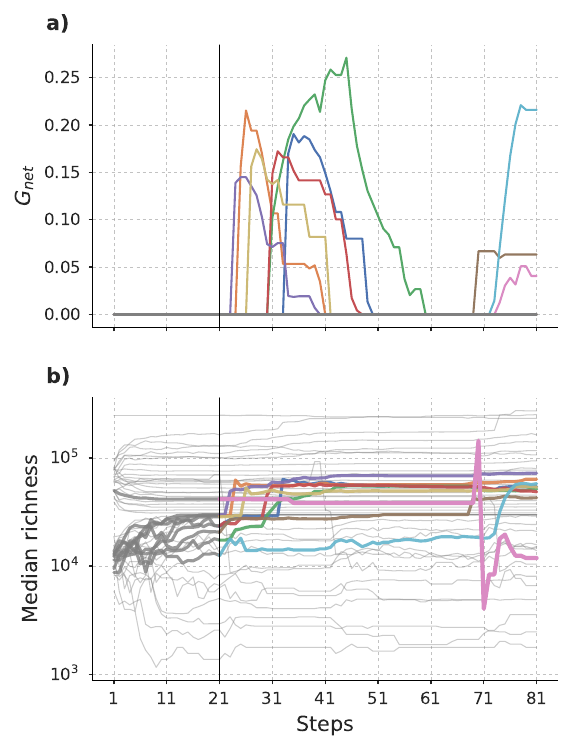}
    \caption{\textbf{$\mathbf{\mathcal{G}_{net}}$ peaks capture richness transitions in neighbourhoods:} \textbf{a)} Time series of $\mathcal{G}_{net}(t,\Delta)$, with fixed values of the time window width $\Delta=15$, of only the nodes that undergo sharp peaks of $\mathcal{G}_{net}$. \textbf{b)} Time series of the richness (median of agents' income) of all nodes in the network: the colored curves correspond to the same nodes whose time series of $\mathcal{G}_{net}$ is presented in \textbf{a)}. The black vertical line in both plots indicates the initial time step when $H$ agents become eligible to relocate, occurring $\theta$ steps after the simulation's start.}
    \label{fig:richness}
\end{figure}

We analyse the relationship between $\mathcal{G}_{net}$ and socioeconomic neighborhood dynamics in Figure \ref{fig:richness}. Nodes exhibiting $\mathcal{G}_{net}$ peaks are colour-coded, while those where $\mathcal{G}_{net}$ remains at zero are depicted in grey (Figure \ref{fig:richness}a). We now consider the median of the income distribtion of agents populating a neighbourhood at a specific time as the \emph{richness} of that neighbourhood at that time. Richness time series reveal a starting trimodal distribution: high-richness central nodes (neighbourhoods), intermediate-richness nodes, and low-richness nodes (Figure \ref{fig:richness}b). Several nodes transition from low to intermediate richness, coinciding with $\mathcal{G}_{net}$ peaks. A notable exception (pink curve) displays a sharp richness increase followed by a steep decline, ultimately transitioning from low to middle richness.  $\mathcal{G}_{net}$ correctly detects gentrification only after the moment this curve transitions from low to middle richness, disregarding the earlier fluctuations. The overall correspondence between the colour-coded curves in both figures validates the capacity of $\mathcal{G}_{net}$ to identify gentrification solely based on relocation patterns.

\newpage
\subsection{Gentrification follows city density.}

\begin{figure}[]
    \centering
    \includegraphics[width=\linewidth]{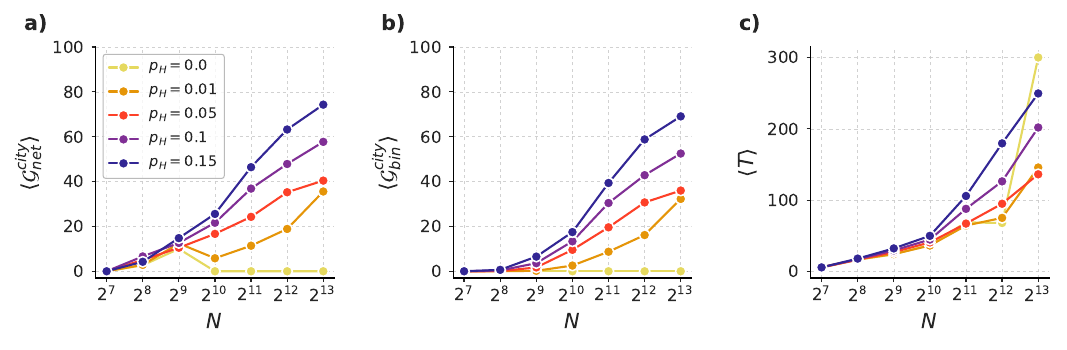}
    \caption{\textbf{City density drives gentrification dynamics:}
 Average results across 150 simulations for varying agent populations (x-axis, logarithmic scale) and high-income agent relocation rates $p_H$ (colors). \textbf{a,b)} City-wise gentrification levels measured by $\mathcal{G}_{net}^{city}$ (a) and $\mathcal{G}_{bin}^{city}$ (b). \textbf{c)} Number of average simulation time steps. Higher urban densities facilitate gentrification and increase convergences time.}
    \label{fig:several_agents}
\end{figure}

In Figures \ref{fig:several_agents}a-c we show the relationship between urban density and gentrification levels across different values of the model parameter $p_H$. We conduct 150 simulations for each configuration, with fixed values of the parameters $\theta = 20$ and $\Delta = 10$. To model increasing urban density, we run each simulation with a different number of agents ($N$) present in the grid, while keeping the grid dimension constant at 7x7 along with the capacity of individual nodes $K$. 

City-wise gentrification levels, as captured by $\mathcal{G}_{net}^{city}$ (Figure \ref{fig:several_agents}a) and $\mathcal{G}_{bin}^{city}$ (Figure \ref{fig:several_agents}b), are characterised by a clear trend: as city population density increases, so does the propensity for gentrification, in terms of number of gentrification events as observed by the two measures (see Methods for details), averaged over the different runs of the model. Furthermore, this effect is amplified by the $H$ agents relocation rate $p_H$, as the curves in the two figures increase monotonically with $N$, with the exception of 0 or very low values of $p_H$ and low $N$ for $\mathcal{G}_{net}$, as shown in Figure \ref{fig:several_agents}a.
In Figure \ref{fig:several_agents}c, we show the relationship between city density and the average convergence time $\langle T\rangle$, the mean number of steps required to reach the termination condition over the 150 simulations, where $L$ agents can no longer move. The average convergence time $\langle T\rangle$ increases with both $N$ and $p_H$, except when $p_H$ = 0. In this case, where $H$ agents are present but stationary and the city is extremely dense, the model does not reach the termination condition within the imposed 300-step limit when $N\geq 2^{13}$. 

 In addition to city-level dynamics, neighbourhood density, limited by fixed capacity, plays an interesting role. In particular, we observe that the initially less populated low-income neighbourhoods become increasingly crowded over time.This is driven by the displacement pressure exerted by higher-income agents, who push lower-income agents out of more desirable areas. As a result, low-income agents concentrate into a few affordable neighborhoods, leading to significant overcrowding. A detailed analysis of this process is provided in Supplementary Information 5.

Overall these results suggest how higher urban densities lead to the emergence of more gentrification waves throughout the evolution of a city.

\section{Discussion}
\label{sec:discussion}

Our study introduces an agent-based model of gentrification that categorises inhabitants of a city into three income groups -- low, medium and high -- and simulates agent movements within a grid-based urban environment driven by socioeconomic factors. The model effectively captures the essence of gentrification dynamics, consisting in the displacement of lower income inhabitants of a neighbourhood of the city caused by a simultaneous inflow of wealthier citizens\cite{glass2013aspects}. This characteristic of gentrification
is evidenced by the results of our simulations and quantitatively described by our proposed network-based measure. 

%Role of high-income agents: 
We find that even a small proportion of high-income agents (5\% of the population) significantly affect the dynamics of gentrification. When high-income agents do not move, even if still present in the model city, no gentrification is detected, and the model does not converge within the imposed limit of 300 steps. However, introducing even a low probability for the movement of high-income agents ($p_H$ = 1\%) leads to model convergence in approximately 150 steps, while 40\% of the city neighbourhoods experience at least one gentrification wave. Interestingly,  when the model is initialized with a random configuration in which agents of all three types are substantially represented in each neighborhood (see Supplementary Figures 10–13), rather than with a core-periphery structure, it rapidly converges to a segregated state. This outcome is equivalent to the final configuration obtained in simulations starting from the core-periphery setup, although it lacks any preserved spatial organization.

%Effectiveness of quantitative measures: 
Our measures, $\mathcal{G}_{count}$ and $\mathcal{G}_{net}$, represent two distinct approaches to quantify gentrification. The count-based measure $\mathcal{G}_{count}$ evaluates the concentration of agents of the three types in each neighborhood over a time window $\Delta$. A significance threshold ($n_{H,M}^*$) is needed to detect when middle- and high-income agents are over-represented and, therefore, define the binarised version of the count-based measure, $\mathcal{G}_{bin}$. Such measure thus captures neighborhood transitions from under- to over-representation of middle- and high-income agents, indicating the completion of gentrification. The network-based measure $\mathcal{G}_{net}$ tracks the net inflow of middle- and high-income agents and the simultaneous outflow of lower-income agents over $\Delta$. This measure is thus rooted in temporal network analysis, where the existence of structures\cite{backbones,span-cores,TRC} in the networks under study is related to the simultaneity of the interactions (edges) between pairs or groups of nodes. Furthermore, this network-based approach avoids more or less arbitrary thresholds and aligns with available commuting flow data\cite{barbosa2018human, noulas2012tale, pappalardo2024survey, mauro2022generating}, which could serve as a proxy given the absence of residential relocation records. $\mathcal{G}_{net}$ helps identifying early signs of gentrification by tracking peaks or plateaus in its trajectory, correlating higher-income resident influx with lower-income displacement. The cross-correlation computed between the count-based and the network-based measures highlights how peaks in $\mathcal{G}_{net}$ are consistently and significantly observed in advance with respect to $\mathcal{G}_{count}$. The analysis of the randomized versions of our model, in Supplementary Information 1, shows that while gentrification events caused by random agent-relocations are still detected, no significant cross-correlation exists between our two measures, emphasizing the importance of the agents' decision-making rules in our model for predicting transitions. Our results hold true also when relaxing the termination condition of our model, thus allowing longer simulations (Supplementary Information 8). Furthermore, the analyses of longer simulations reveal a quasi-periodic occurrence of gentrification \emph{waves}, defined as peaks of the time series of the volume of gentrification events unfolding across the whole city. These results demonstrate that our minimal model captures the essential features needed to reproduce gentrification: a heavy-tailed income distribution, few income classes, and distinct relocation strategies -- notably the profit-driven behavior of high-income agents. Moreover, our network-based measure $\mathcal{G}_{net}$ enables earlier detection of gentrification compared to count-based metrics, potentially aiding policymakers in preventing low-income displacement.

%Limitations and Future Directions

While our model provides valuable insights into gentrification dynamics, it has limitations that future research could address. The constant population size and static income-group assignments could be expanded to incorporate population growth and inter-city migration flows, potentially using a network-based approach to disentangle endogenous and exogenous causes of gentrification\cite{rotterdam}. With the implementation of these extensions of our model, a complementary measure of \emph{impoverishment} of a neighbourhood, such as the one we propose in the Supplementary Material (Supplementary Material 7) could help elucidate, in a quantitative manner, the causal relationships between the periods of impoverishment/disinvestment and gentrification of neighbourhoods. Multiple property ownership per agent could be introduced to model wealth concentration and short-term rental effects. The grid-based urban representation could be enhanced with more complex geographical features, although our results hold for both 7$\times$7 and 9$\times$9 grids (Supplementary Information 3). Moreover, simulating policy interventions\cite{mauro2024role} could provide insights for urban planners, particularly regarding density restrictions given our findings on city density and gentrification probability. 
%Finally, future research

In conclusion, our agent-based model and novel quantitative measures offer a powerful framework for understanding and predicting gentrification processes. This approach not only advances our theoretical understanding of gentrification but also provides quantitative what-if tools for early detection and potential mitigation of its effects in real-world urban environments.

\section{Methods}
\label{sec:methods}
\subsection{Agents' income.} 
 At the beginning of a simulation, each agent is assigned a fixed income $w$, based on data from the 2022 USA Social Security Administration report\cite{ssa2022compensation}. The assignment process uses the income brackets and population percentages provided in this report. Each agent is assigned to an income bracket with probability proportional to the US population within that bracket, and then the agent's specific income $w$ is randomly selected from within their assigned bracket.
Agents are categorized into three groups based on their assigned incomes: $L$ (low-income) agents with incomes up to \$29,999.99, encompassing the 2022 poverty line for a family of four (\$27,750);  $H$ (high-income) agents representing the top 5\% of earners; and $M$ (middle-income) agents comprising all remaining individuals.

\subsection{Gentrification: peaks, shifts and aggregate measure} 
A cell $i$ undergoes gentrification events according to $\mathcal{G}_{bin}$ at all times $t$ where there is a binary \textit{shift} from 0 to 1 in the time series throughout the simulation:
\begin{equation}
T_{shift}^i = \{t : \mathcal{G}_{bin}^i(t,\Delta) = 1 \wedge \mathcal{G}_{bin}^i(t-1,\Delta) = 0\}.
\end{equation}
A cell $i$ undergoes gentrification events according to $\mathcal{G}_{net}$ at all times $t$ where there is a peak in its time series throughout the simulation. A peak is defined as either a local maximum or the start of a plateau after a growing phase:
\begin{equation}
T_{peak}^i = \left\{t : \left[
\begin{array}{c}
(\mathcal{G}_{net}^i(t,\Delta) > \mathcal{G}_{net}^i(t-1,\Delta) \wedge \mathcal{G}_{net}^i(t,\Delta) > \mathcal{G}_{net}^i(t+1,\Delta)) \;  \\
\vee
\\
(\mathcal{G}_{net}^i(t,\Delta) > \mathcal{G}_{net}^i(t-1,\Delta) \wedge \mathcal{G}_{net}^i(t,\Delta) = \mathcal{G}_{net}^i(t+1,\Delta))
\end{array}
\right]\right\}.
\end{equation}
We define the gentrification level of the city as the percentage of cells of the city that experience at least one gentrification event according to each measure:
\begin{equation}
\mathcal{G}_{bin}^{city} = \frac{\sum_{i=1}^{N_{\text{cells}}} \mathbf{1}(|T_{shift}^i| > 0)}{N_{\text{cells}}} \times 100
\quad,\quad
\mathcal{G}_{net}^{city} = \frac{\sum_{i=1}^{N_{\text{cells}}} \mathbf{1}(|T_{peak}^i| > 0)}{N_{\text{cells}}} \times 100,
\end{equation}
where $\mathbf{1}(\cdot)$ is the indicator function.

\subsection{Lagged cross-correlation}
We compute the mean cross-correlation $\langle R\rangle$ by calculating the lagged cross-correlation between pairs of time series of $\mathcal{G}^i_{net}(t,\Delta)$ and $\mathcal{G}^i_{bin}(t,\Delta)$ corresponding to each cell on the grid-view of a model city, for several values of the lag $\tau$, and then averaging over all cells. To compute the cross-correlation $R^i$ between the two time series for a cell $i$, we transform the two time series into two binary vectors (see Suplementary Information 4 for further details) $\widetilde{\mathcal{G}}_{net}^i$ and $\widetilde{\mathcal{G}}_{bin}^i$ of length $T$, where the $t_k$-th entry is $1$ if the original corresponding time series has a peak or 0-1 transition at time $t_k\in[0,T]$, respectively:
\begin{equation}
    (\widetilde{\mathcal{G}}_{net}^{i})_{t_k}=\begin{cases}
    1, & \text{if $t_k \in T_{peak}^i$},\\
    0, & \text{otherwise}.
  \end{cases}
\end{equation}
\begin{equation}
      (\widetilde{\mathcal{G}}_{bin}^{i})_{t_k}=\begin{cases}
    1, & \text{if $t_k \in T_{shift}^i$},\\
    0, & \text{otherwise}.
  \end{cases}
\end{equation}
For each value of the lag $\tau\in[-15,+15]$, we compute the lagged cross-correlation between the two vectors and obtain a value of $R^i(\tau)$ for each cell $i$ on the grid. We then calculate $\langle R\rangle$ by averaging, for each value of the lag $\tau$, the values of $R^i(\tau)$ over all cells $i\in[0,N_{\text{cells}}]$:
\begin{equation}
    \langle R(\tau)\rangle = \frac{\sum_{i \in[0,N_{\text{cells}}]}R^i(\tau)}{N_{\text{cells}}}.
\end{equation}
To establish a baseline for comparison, we generate a null distribution by computing, for each cell $i$, the cross-correlation $R$ between $50$ pairs of randomly reshuffled versions of the two vectors $\widetilde{\mathcal{G}}_{net}^i$ and $\widetilde{\mathcal{G}}_{bin}^i$.

\section*{Code availability statement}
The code for the implementation of our model and to reproduce our analyses can be found at 
\url{https://github.com/mauruscz/Gentrification}.

\section*{Acknowledgments}

We thank Daniele Fadda for its precious support with the visualizations.
We thank Dino Pedreschi, Timothy LaRock, Rohit Sahasrabuddhe, Andrea Beretta, Giuliano Cornacchia, Margherita Lalli, Daniele Gambetta, Emanuele Ferragina and Salvatore Citraro for their useful suggestions.

This work has been partially supported by: EU project H2020 SoBigData++ G.A. 871042; PNRR (Piano Nazionale di Ripresa e Resilienza) in the context of the research program 20224CZ5X4 PE6 PRIN 2022 ``URBAI – Urban Artificial Intelligence" (CUP B53D23012770006), funded by the European Commission under the Next Generation EU programme; and by PNRR - M4C2 - Investimento 1.3, Partenariato Esteso PE00000013 - ``FAIR – Future Artificial Intelligence Research" – Spoke 1 ``Human-centered AI", funded by the European Commission under the NextGeneration EU programme; project ``SoBigData.it - Strengthening the Italian RI for Social Mining and Big Data Analytics", prot. IR0000013, avviso n. 3264 on 28/12/2021.\\
NP and RL received funding from EPSRC Grant Ref EP/V013068/1.

\section*{Author contributions} 
GM: study conceptualisation, model implementation, experiment design, execution of experiments, code implementation, interpretation of results, writing, plots and images, study management. 
NP: study conceptualisation, model implementation, experiment design, interpretation of results, writing, plots design.
LP: study conceptualisation, experiment design, interpretation of results, writing, study direction. 
RL: study conceptualisation, experiment design, interpretation of results, writing, study direction.
All authors read and approved the final manuscript.

\section*{Competing interests} The authors declare no competing interests.

\section*{Corresponding authors} 
Giovanni Mauro - giovanni.mauro@sns.it

\enlargethispage*{13pt}

\section*{ORCID}

\noindent Giovanni Mauro - \url{https://orcid.org/0000-0001-8067-984X}

\noindent Nicola Pedreschi - \url{https://orcid.org/0000-0003-1582-6246}

\noindent Renaud Lambiotte - \url{https://orcid.org/0000-0002-0583-4595}

\noindent Luca Pappalardo - \url{https://orcid.org/0000-0002-1547-6007}

\bibliographystyle{ws-acs}
\bibliography{ws-acs}

\end{document}

\title{Dynamic models of Gentrification\\ Supplementary Information}
\author{Giovanni Mauro, Nicola Pedreschi, Renaud Lambiotte, Luca Pappalardo}
\date{}
\maketitle
\tableofcontents

\section{Comparison with null models}

We devised two null models to validate our findings. In the first version ($R_1$), agents behave as pure random walkers:
\begin{itemize}
\item At each time step, every agent decides whether to move with the same, fixed probability (e.g., 50\% chance of moving).
\item If an agent decides to move, it relocates to a cell with space chosen with a uniform random probability throughout the grid.
\end{itemize}

In the second version ($R_2$), agents follow the same rules for deciding whether to relocate as in the main model:
\begin{itemize}
\item Low-income agents move when their economic condition falls below a critical threshold
\item Middle-income agents relocate when experiencing either extreme poverty or wealth
\item High-income agents' mobility is governed by the parameter $p_H$
\item However, like in the first null model, if an agent decides to move, they relocate to a randomly chosen available cell
\end{itemize}

Both models remove the targeted relocation aspect, with $R_1$ additionally removing the evaluation criteria, thus providing two distinct baselines for comparison with our main model.

\begin{figure}[H]
    \centering
    \includegraphics[width=\linewidth]{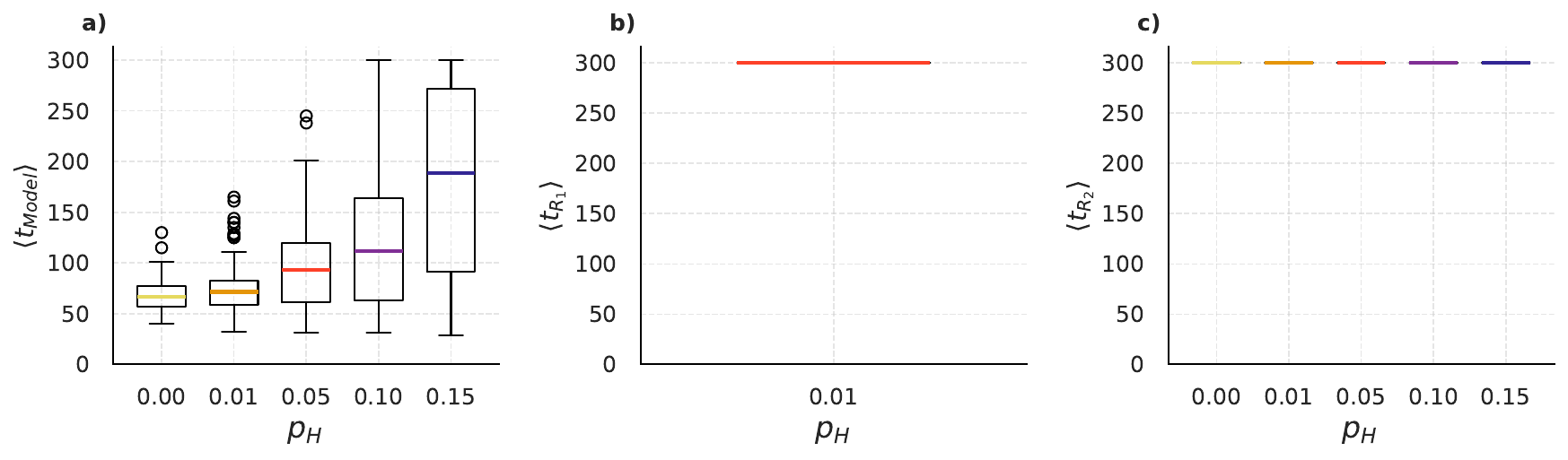}
    \caption{Average convergence time comparison for the three models. Both null models never reach convergence, while for the main model the convergence time grows with the $p_H$ parameter.}
    \label{fig:SUPPL-times}
\end{figure}

Figure \ref{fig:SUPPL-times} compares convergence times across the three model variants- While the original model eventually reaches convergence with times increasing as $p_H$ increases, both randomized variants fail to converge at all, as they always reach the 300 steps limit. This systematic difference suggests that random relocation, whether in decision-making or destination choice, prevents the system from reaching stable configurations, highlighting the importance of targeted movement in our main model.

\begin{figure}[H]
    \centering
    \begin{subfigure}[b]{0.45\textwidth}
        \centering
        \includegraphics[width=\linewidth]{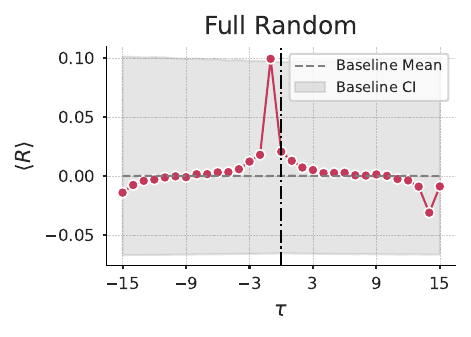}
        \label{fig:random}
    \end{subfigure}
    \hfill
    \begin{subfigure}[b]{0.45\textwidth}
        \centering
        \includegraphics[width=\linewidth]{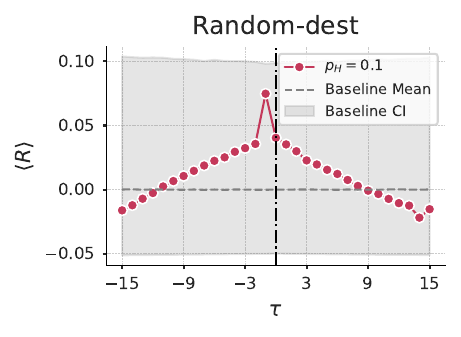}
        \label{fig:randomdest}
    \end{subfigure}
    \caption{Cross-correlation analysis for the two null models. Both models show comparable patterns, further validating that random destination choice cannot explain the empirical observations.}
    \label{fig:nullmodels}
\end{figure}

We applied the same statistical analysis used in the main text to assess the statistical relevance of $\mathcal{G}_{net}$ anticipating $\mathcal{G}_{bin}$ for these baseline models. The results are presented in Figure \ref{fig:nullmodels}. Two key observations emerge:
\begin{enumerate}
\item Neither null model produces statistically significant results, indicating an absence of observable anticipatory patterns regardless of whether the movement decision is random or follows socioeconomic rules.
\item In both cases, the peak of the cross-correlation, although not significant, occurs at $\tau = -1$. This is an expected outcome, as the $\mathcal{G}_{net}$ measure is based on flow data, while $\mathcal{G}_{bin}$ is derived from count data. Specifically, the flow of agents moving at one time step ($\mathcal{G}_{net}$) is reflected in the count of agents ($\mathcal{G}_{bin}$) in the subsequent time step, naturally creating a lag of -1.

\end{enumerate}

The absence of statistically significant anticipatory patterns in both null models reinforces the validity of our main model's findings. It suggests that the observed anticipatory behaviour is not a random artefact but rather an emergent property arising from the interplay of our model's evaluation criteria and, crucially, the targeted relocation strategies. This comparison provides strong evidence for the robustness of our main model and the significance of its results in capturing complex socio-economic dynamics.

\section{Early warning consistency across parameters and measures}
In the main text, we presented the results of the cross-correlation analysis for a single value of $p_H$, the probability of high-income agents relocating. Here, we demonstrate that the observed anticipatory patterns are robust across different values of $p_H$ and alternative analytical measures.

\begin{figure}[H]
    \centering
    \includegraphics[width=1\linewidth]{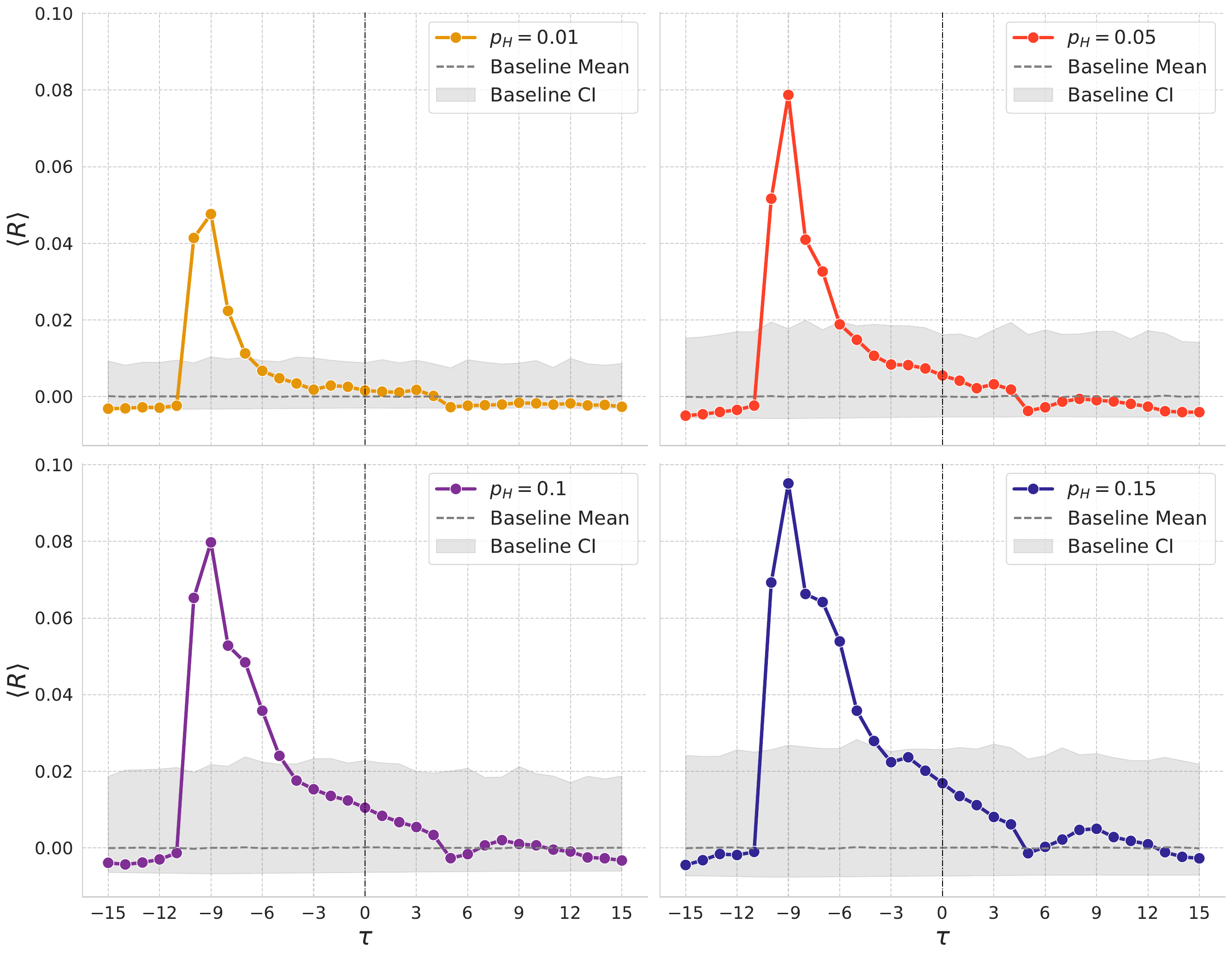}
    \caption{Cross-correlation analysis between $\mathcal{G}_{net}$ and $\mathcal{G}_{bin}$ for various $p_H$ values. Each panel represents a different $p_H$. The x-axis shows the lag $\tau$, and the y-axis represents the cross-correlation coefficient. Shaded areas indicate 95\% confidence intervals. The consistent peak at negative lags across all $p_H$ values demonstrates the robustness of the anticipatory relationship between $\mathcal{G}_{net}$ and $\mathcal{G}_{bin}$, regardless of the probability of high-income agent relocation.}
    \label{suppl-fig:cross-all}
\end{figure}

Figure \ref{suppl-fig:cross-all} illustrates the consistency of the cross-correlation analysis results for various $p_H$ values. The anticipatory relationship between $\mathcal{G}_{net}$ and $\mathcal{G}_{bin}$ remains stable across different probabilities of high-income agent relocation, indicating that this phenomenon is not sensitive to specific parameter choices within our model.

\begin{figure}[H]
    \centering
    \includegraphics[width=1\linewidth]{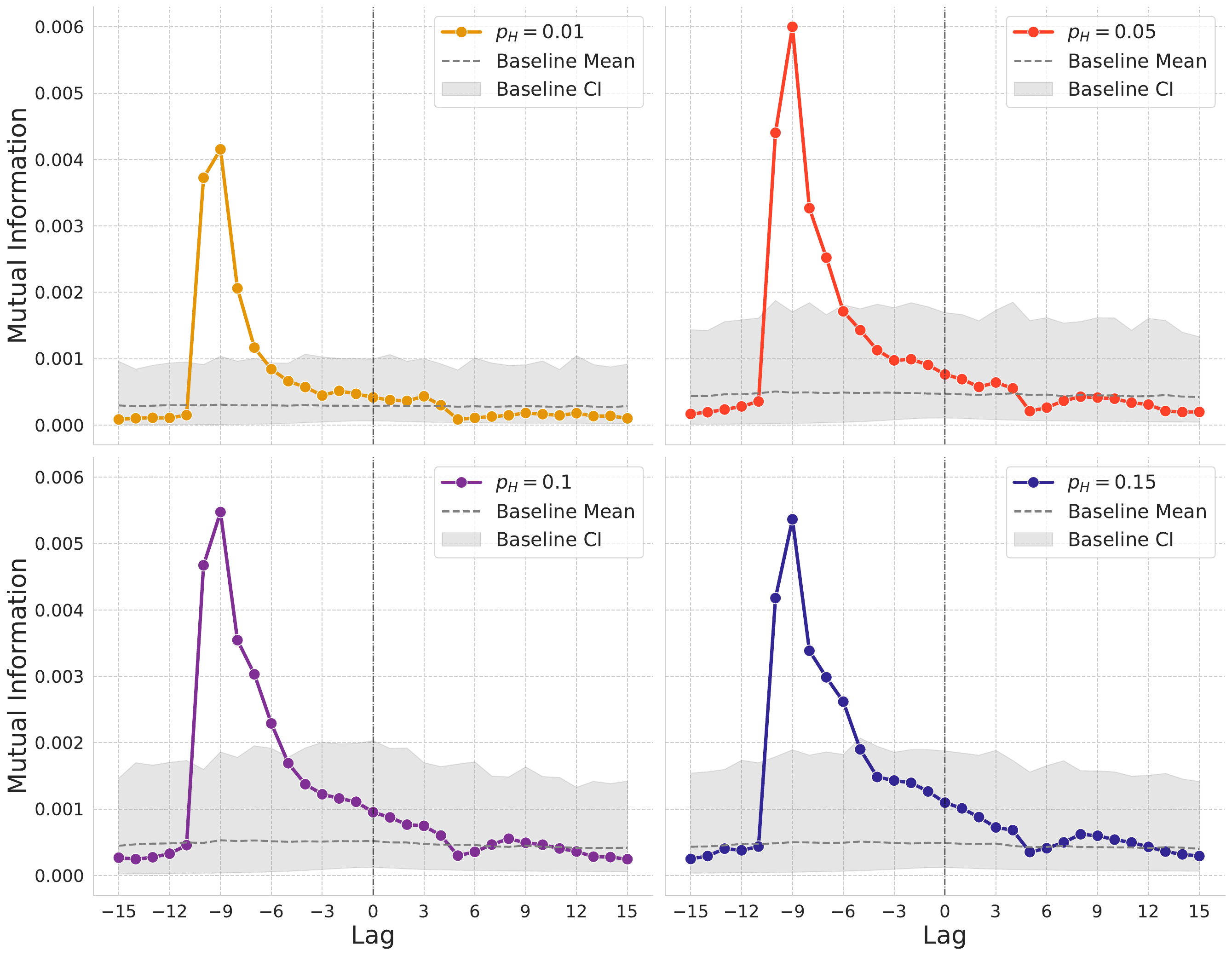}
    \caption{Mutual Information (MI) analysis between $\mathcal{G}_{net}$ and $\mathcal{G}_{bin}$ for various $p_H$ values. Each panel corresponds to a different $p_H$. The x-axis represents the lag $\tau$, while the y-axis shows the MI value. Shaded areas indicate 95\% confidence intervals. The patterns observed are consistent with the cross-correlation analysis, exhibiting peaks at negative lags. Note that MI values are approximately one order of magnitude smaller than cross-correlation coefficients, which is expected due to the different nature of these measures. The consistency across $p_H$ values further supports the robustness of the anticipatory relationship between $\mathcal{G}_{net}$ and $\mathcal{G}_{bin}$.}
    \label{suppl-fig:mutualinfo-all}
\end{figure}

To further validate our findings, we conducted an additional analysis using Mutual Information (MI) instead of cross-correlation, as shown in Figure \ref{suppl-fig:mutualinfo-all}. MI provides a more general measure of statistical dependence, capturing both linear and non-linear relationships between variables. The results from this analysis corroborate our cross-correlation findings, exhibiting similar anticipatory patterns. It's worth noting that while the overall trends are consistent, the absolute values of MI are approximately one order of magnitude smaller than those of the cross-correlation analysis.

The consistency of results across different $p_H$ values and analytical methods reinforces the robustness of our findings, suggesting that the anticipatory patterns observed in our model are a fundamental feature of the gentrification dynamics we've simulated, rather than an artifact of specific parameter choices or analytical approaches.

\section{Results are Consistent for a Larger (9x9) City}

To assess the scalability and robustness of our findings, we extended our analysis to a larger urban environment, specifically a 9x9 grid city. This expansion allows us to verify whether the patterns observed in our original model persist in a more complex urban setting with a greater number of neighborhoods and potential interactions.

\begin{figure}[H]
    \centering
    \includegraphics[width=1\linewidth]{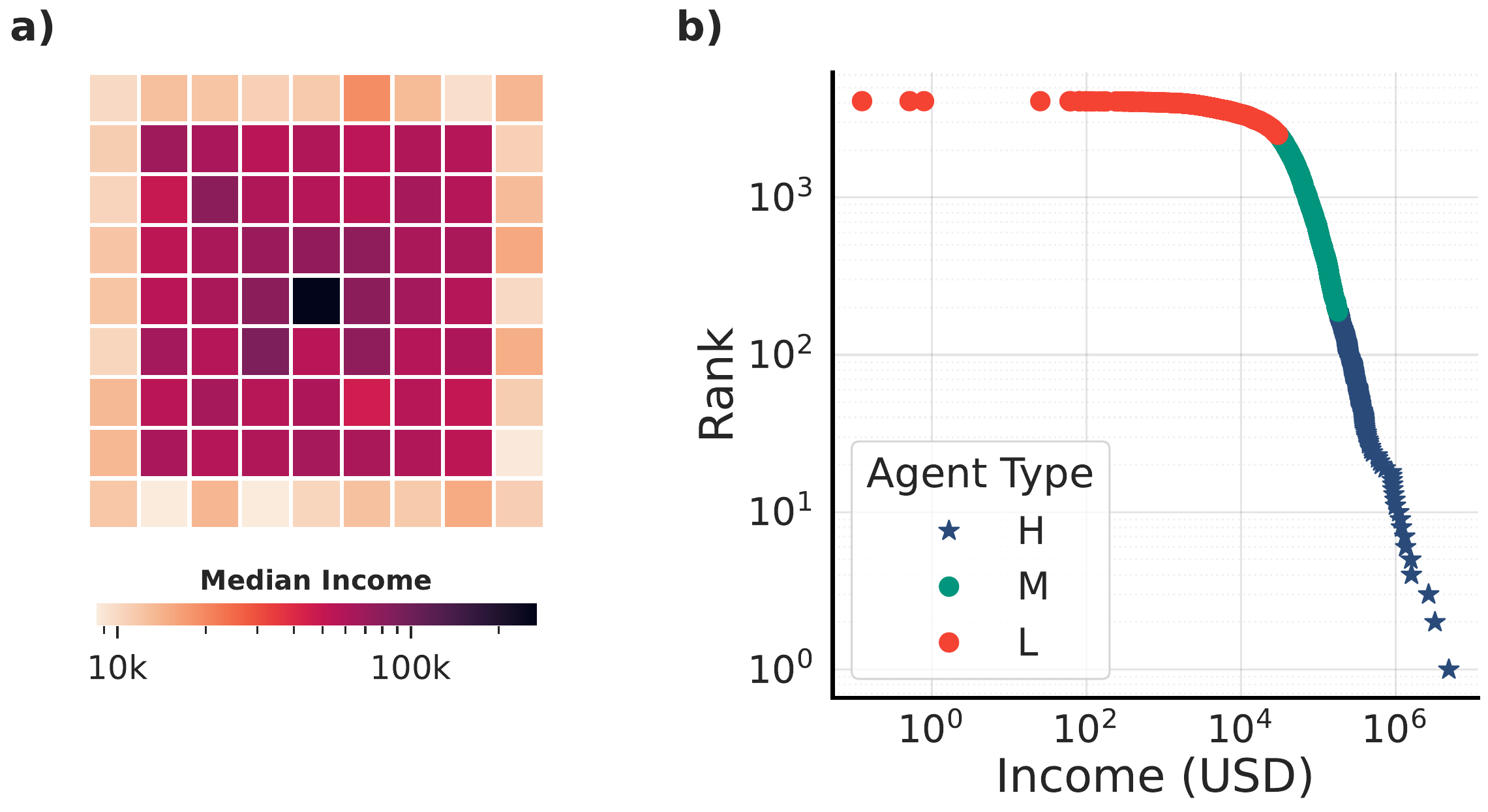}
    \caption{\textbf{Income distribution and spatial arrangement in a 9x9 city model.} \textbf{Left panel:} Heatmap depicting the median income of each cell at the initial state of the simulation for a 9x9 grid city. Darker colors indicate higher median incomes. The hierarchical structure and spatial clustering of income levels are preserved in this larger model. \textbf{Right panel:} log-log plot of the initial distribution of agent incomes sampled from the 2022 USA Social Security Administration report. N = $2^{12}$ agents. }
    \label{suppl-fig:big-heatmap-income}
\end{figure}

Figure \ref{suppl-fig:big-heatmap-income} presents the spatial arrangement and income distribution for this larger city model.

\begin{figure}[H]
    \centering
    \includegraphics[width=\linewidth]{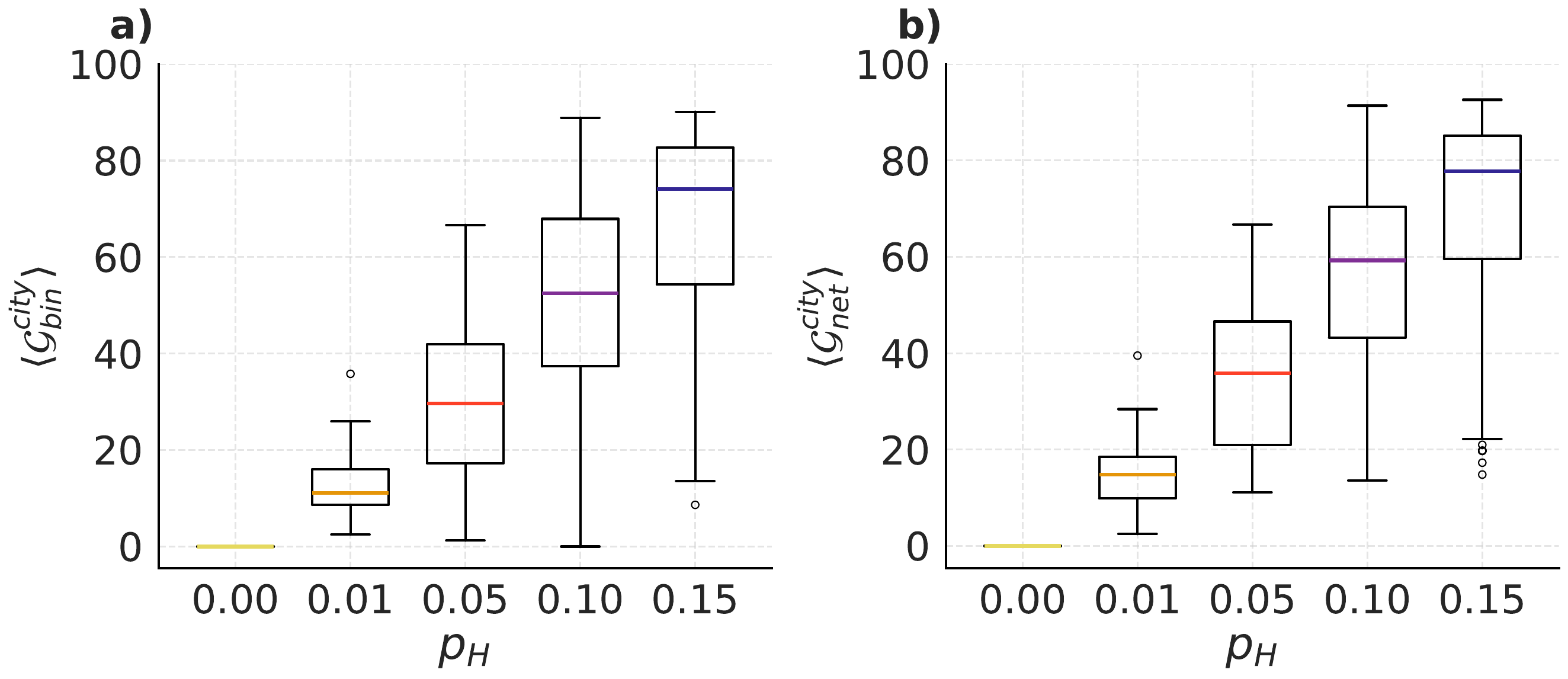}
    \caption{\textbf{Gentrification dynamics in a 9x9 city model.} We show gentrification levels and times across 150 simulations with fixed $\epsilon=20$, $N=2^{12}$, and varying values of $p_H$ (probability of high-income agent relocation). The left panel shows gentrification levels according to $\mathcal{G}_{bin}^{city}$, while the right panel displays levels according to the network-based measure $\mathcal{G}_{net}^{city}$. The x-axis represents different $p_H$ values, and the y-axis shows the percentage of gentrified cells. Box plots display the distribution of gentrification levels across simulations. The patterns observed closely resemble those in the smaller model: absence of gentrification at $p_H=0$, abrupt transition to gentrification at low $p_H$ values, and monotonic increase in gentrification levels as $p_H$ increases. }
    \label{suppl-fig:big-boxes}
\end{figure}

Figure \ref{suppl-fig:big-boxes} presents the results of simulations conducted on the 9x9 city grid, mirroring the analysis shown in Figure 3 of the main text. Remarkably, the gentrification patterns observed in this larger model closely resemble those of the original smaller model, providing strong evidence for the scalability and consistency of our findings.

In the 9x9 city model, we observe that the critical role of $H$ agent mobility in initiating gentrification is preserved. At $p_H=0$, both $\mathcal{G}_{bin}^{city}$ and $\mathcal{G}_{net}^{city}$ show 0\% gentrification, indicating a complete absence of the phenomenon when high-income agents are static. The abrupt transition to gentrification with minimal $H$ agent mobility ($p_H=0.01$) is also evident in the larger model, with both metrics showing gentrification in a significant proportion of cells.
As $p_H$ increases, we see a similar monotonic rise in gentrification levels, with $\mathcal{G}_{bin}^{city}$ and $\mathcal{G}_{net}^{city}$ exhibiting comparable trends but slightly different magnitudes, consistent with the original model.

\begin{figure}[H]
    \centering
    \includegraphics[width=\linewidth]{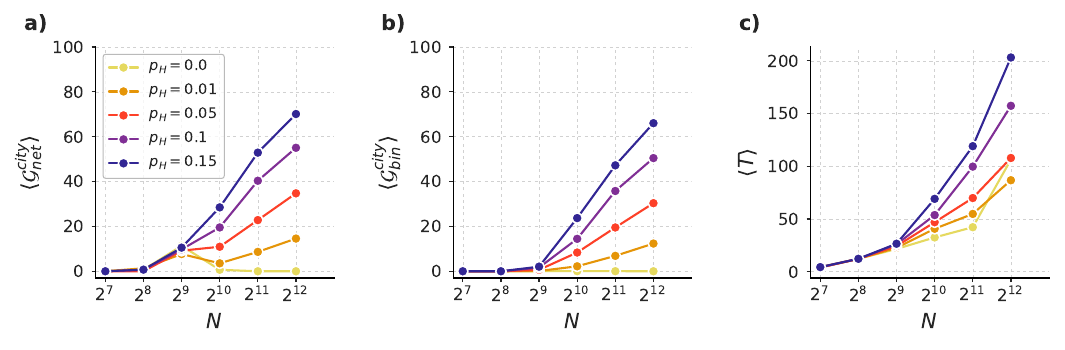}
    \caption{\textbf{Urban density and gentrification dynamics in a 9x9 city model.} We show average results across 150 simulations for varying agent populations (x-axis, logarithmic scale) and high-income agent relocation rates $p_H$ (colors). \textbf{Left panel:} City-wise gentrification levels measured by $\mathcal{G}_{net}^{city}$. \textbf{Middle panel:} City-wise gentrification levels measured by $\mathcal{G}_{bin}^{city}$. \textbf{Right panel:} Average number of simulation time steps (convergence time $\langle T\rangle$). The patterns observed closely resemble those in the smaller 7x7 model: (1) gentrification levels increase with urban density for both measures, (2) this effect is amplified by higher $p_H$ values, and (3) convergence time generally increases with both density and $p_H$.}
    \label{suppl-fig:big-several_agents}
\end{figure}

Figure \ref{suppl-fig:big-several_agents} presents the results of simulations conducted on the 9x9 city grid, mirroring the analysis shown in Figure 6 of the main text. The gentrification patterns observed in this larger model closely resemble those of the original smaller model, providing strong evidence for the scalability and consistency of our findings.

In the bigger city model, we observe that the relationship between urban density and gentrification levels remains consistent across different values of the model parameter $p_H$. As in the smaller model, city-wise gentrification levels, captured by both $\mathcal{G}_{net}^{city}$ and $\mathcal{G}_{bin}^{city}$, show a clear trend: as city population density increases, so does the propensity for gentrification. This effect is amplified by the $H$ agents' relocation rate $p_H$, with the curves increasing monotonically with $N$ for most values of $p_H$.

The relationship between city density and the average convergence time $\langle T\rangle$ in the larger model also mirrors the findings from the original model. The average convergence time increases with both $N$ and $p_H$, except when $p_H$ = 0. In extremely dense scenarios with stationary $H$ agents, the model may not reach the termination condition within the imposed step limit, consistent with the behavior observed in the smaller model.

These observations suggest that the fundamental dynamics of our model, including the processes driving gentrification are not artifacts of the specific grid size used in the main text. Instead, they appear to be scalable properties that emerge from the underlying mechanisms of our agent-based model.

%\bibliographystyle{naturemag}
%\bibliography{biblio}

\section{Binary Vector Transformation for Cross-Correlation Analysis}

\begin{figure}[H]
    \centering
    \includegraphics[width=.5\linewidth]{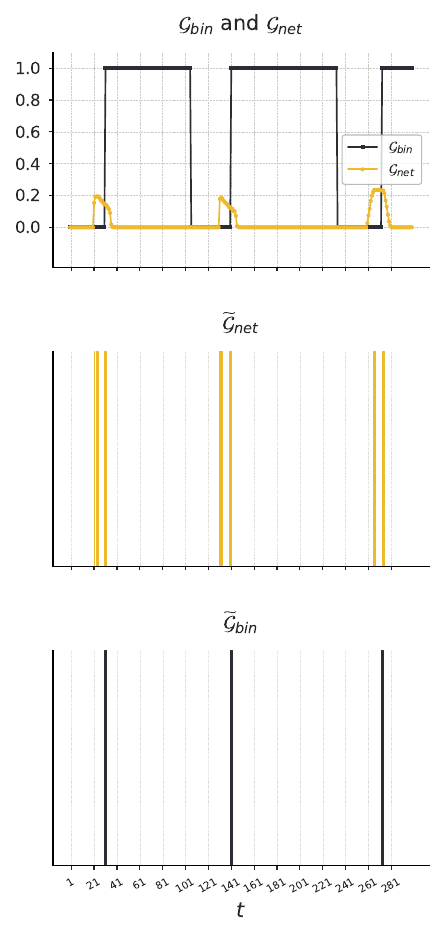}
    \caption{Example of barcode}
    \label{fig:barcodes}
\end{figure}

In our analysis, we compute the cross-correlation $R^i$ between two time series for a cell $i$. To facilitate this process, we transform both time series into binary "barcodes" that represent significant changes or peaks in the original signals. This section details the process of converting the original time series into their corresponding binary vectors, as illustrated in Figure \ref{fig:barcodes}.

The upper part of Figure \ref{fig:barcodes} displays two example time series, denoted as $\mathcal{G}_{net}$ and $\mathcal{G}_{bin}$. We identify key events in these time series: peaks in the continuous signal $\mathcal{G}_{net}$ and transitions in the binary signal $\mathcal{G}_{bin}$. These events are then encoded as binary vectors, $\widetilde{\mathcal{G}}_{net}^i$ and $\widetilde{\mathcal{G}}_{bin}^i$, representing the time points at which the respective events occur. Specifically:

\begin{itemize}
    \item \textbf{For $\widetilde{\mathcal{G}}_{net}^i$:}
    A peak detection algorithm is applied to the time series $\mathcal{G}_{net}$. If a peak is detected at a specific time $t_k$, we assign $(\widetilde{\mathcal{G}}_{net}^i)_{t_k} = 1$; otherwise, $(\widetilde{\mathcal{G}}_{net}^i)_{t_k} = 0$. Thus, the binary vector $\widetilde{\mathcal{G}}_{net}^i$ captures the occurrence of peaks in $\mathcal{G}_{net}$ over the time period $T$.
    
    \item \textbf{For $\widetilde{\mathcal{G}}_{bin}^i$:}
    Similarly, we track the 0-1 transitions in the binary signal $\mathcal{G}_{bin}$. If a transition from 0 to 1 occurs at time $t_k$, we assign $(\widetilde{\mathcal{G}}_{bin}^i)_{t_k} = 1$; otherwise, $(\widetilde{\mathcal{G}}_{bin}^i)_{t_k} = 0$. This forms the binary vector $\widetilde{\mathcal{G}}_{bin}^i$, encoding significant shifts in $\mathcal{G}_{bin}$.
\end{itemize}

The lower part of Figure \ref{fig:barcodes} illustrates the resulting binary vectors, $\widetilde{\mathcal{G}}_{net}^i$ and $\widetilde{\mathcal{G}}_{bin}^i$, which are derived from the original time series. These binary representations, also referred to as "barcodes", simplify the identification of coinciding events (peaks and transitions) between the two time series, making it easier to compute the cross-correlation $R^i$.

It is important to note that in our analysis, the "left" time series (i.e., the series corresponding to negative lags) represents $\mathcal{G}_{net}$, while the "right" time series (i.e., the series corresponding to positive lags) represents $\mathcal{G}_{bin}$.

\section{Impact of neighbourhood density}

\begin{figure}[H]
    \centering
    \includegraphics[width=0.8\linewidth]{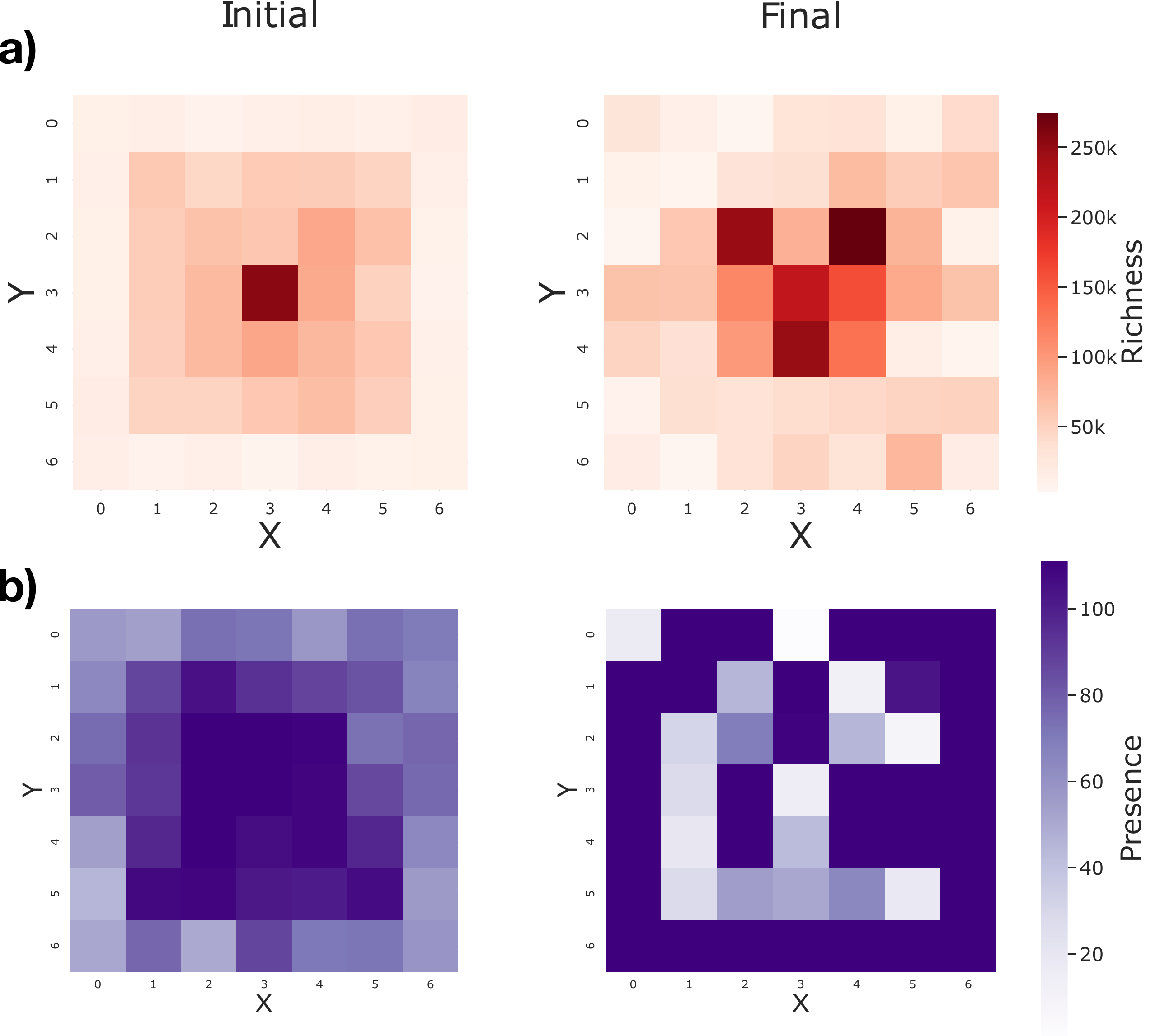}
    \caption{\textbf{Impact of neighbourhood capacity on agents flows:} \textbf{a)} Heatmaps of initial (left) ad final (right) configuration of median income, i.e., richness of neighbourhoods. \textbf{b)} Heatmaps of initial (left) ad final (right) configuration of presence, i.e., fraction of capacity K fulfilled in each neighbourhood, of all neighbourhoods.}
    \label{fig:presence_heatmaps}
\end{figure}

We note from Figure \ref{fig:presence_heatmaps} how the heatmaps of neighbourhoods richness and agents presence, or filling factor, i.e. the fraction of the full capacity K of each neighbourhood that is occupied by agents, are positively correlated at the beginning of the simulation (left) and, instead, they are complementary at the end of the simulation (left). In fact, while the most central neighbourhoods are both the richest and the most densely populated at the beginning of the simulation, the gradual process of \emph{conquering} new neighbourhoods of M and H agents leads to a segregation of all L agents to the most external neighbourhoods (further from the centre) that now become the poorest ( Figure \ref{fig:presence_heatmaps}.a, right) and the most densely populated (Figure \ref{fig:presence_heatmaps}.b, right). 

\section{Comparison of random versus core-periphery initial spatial distribution of agents}

\begin{figure}
    \centering
    \includegraphics[width=0.5\linewidth]{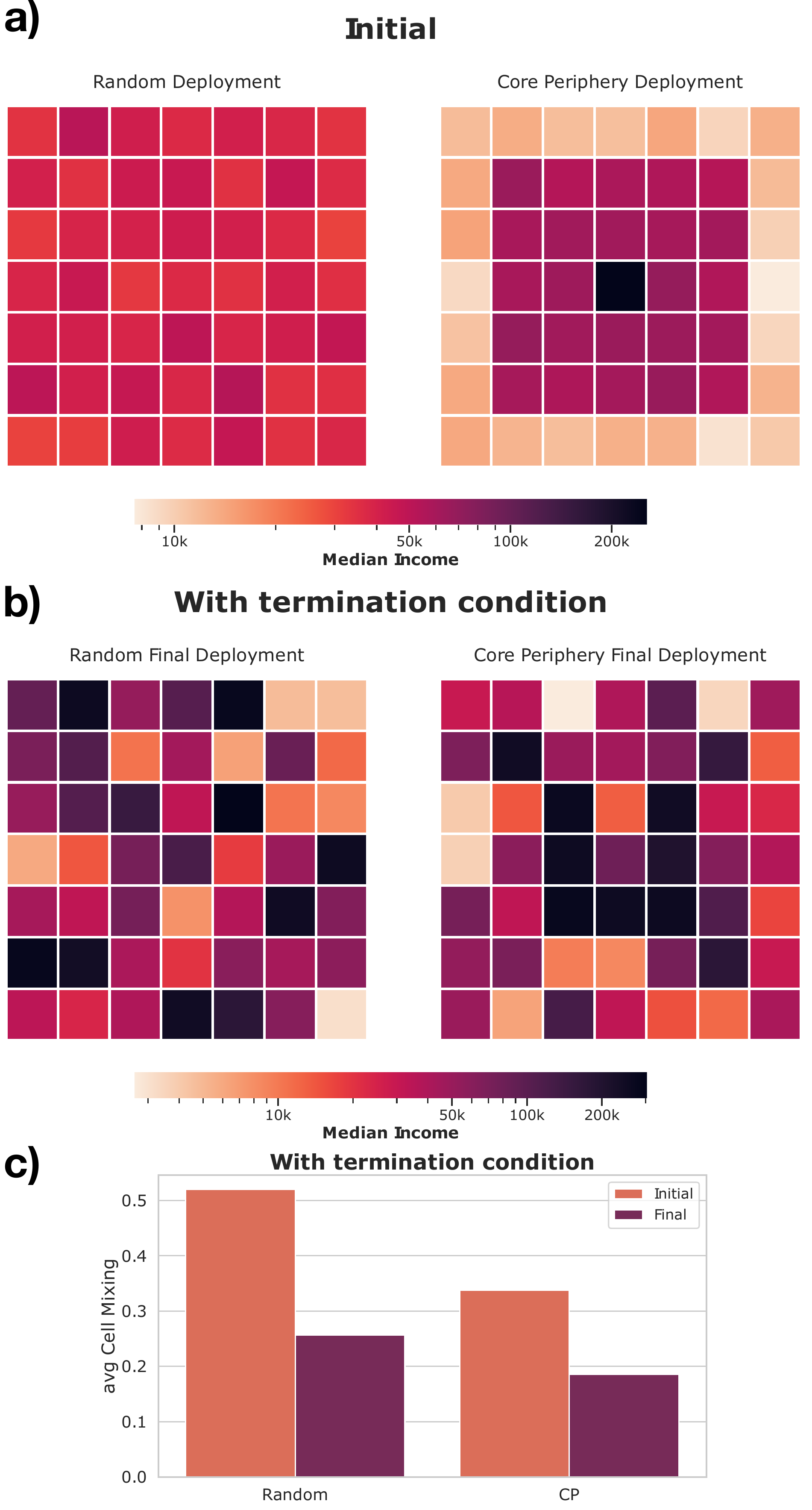}
    \caption{\textbf{Random vs Core-Periphery initial configuration with termination condition:} \textbf{a)} heatmaps of neighbourhood median income at the beginning of the simulation with a random placement of agents, i.e., each neighbourhood has a fraction of agents of the three types close to the fraction of agents of each type in the whole city-grid (left) and of the mono-centric, core-periphery configuration of the city as discussed in the main text (right). \textbf{b)} same heatmaps but obtained from the last step of the simulation of the model, with the termination condition ending the simulation when L agents cannot move. \textbf{c)} On the y-axis is a measure of the average mixing of agents in each neighbourhood, i.e., the Gini index computed over the distribution of income within each grid-cell, averaged over all cells. The orange bars correspond to the value computed for the initial configuration (both for random or core-periphery spatial distribution of agents), while the purple bar corresponds to the same value computed at the last step of the corresponding simulation.}
    \label{fig:mixing_termination}
\end{figure}

In Figure \ref{fig:mixing_termination} we note how the random initial configuration is clearly the most mixed, while the core-periphery configuration is, by design, the most segregated (Figure \ref{fig:mixing_termination}.a). However, at the end of both simulations, i.e., when all L agents are unable to move, the two grids are very similar: while the heatmap of neighbourhood richness at the end of the simulation of the core-periphery structure, seems to still have some spatial organization, both have a minority of cells with high richness, and a majority of poorer cells. The similarity between these two final conditions is quantified by computing the Gini index of the distribution of income of agents in each cell, and then averaging such value over all cells, thus yielding a city-wise level of income-mixing within a neighbouhood (Figure \ref{fig:mixing_termination}.c). While the mixing is, obviously, the highest in the initial random configuration, both simulations converge to a final configuration that is almost equally segregated.

\begin{figure}[H]
    \centering
    \includegraphics[width=0.7\linewidth]{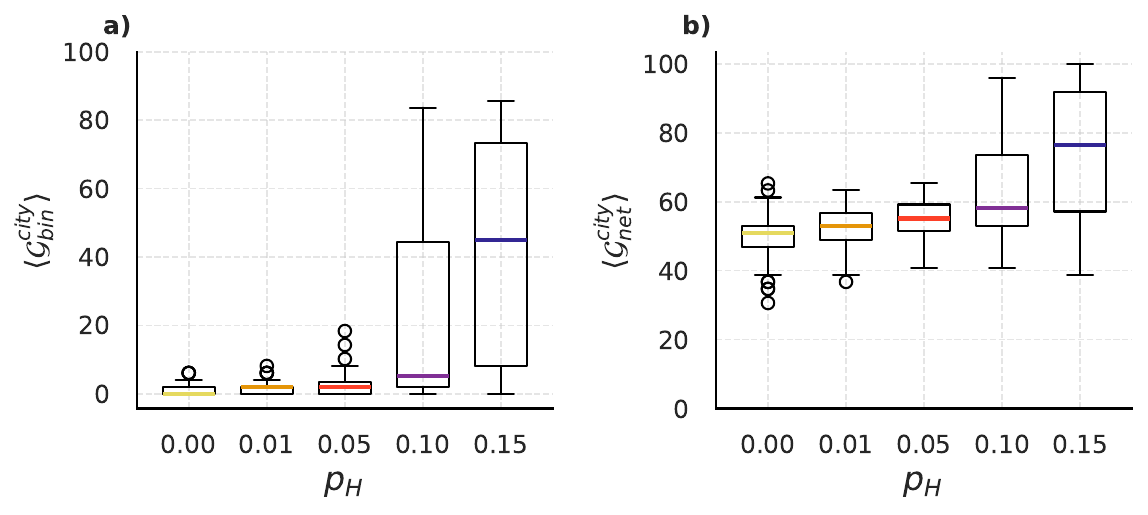}
    \caption{\textbf{City-wise gentrification in the random initial configuration:} Gentrification level across 150 simulation of the model with random initial spatial distribution of agents, a fixed $\theta=20$, $N=2^{12}$ and varying value of $p_H$ according to $\mathcal{G}_{bin}^{city}$ (\textbf{a}) and to the network based measure $\mathcal{G}_{net}^{city}$ (\textbf{b}).}
    \label{fig:random_boxplots}
\end{figure}

\begin{figure}[H]
    \centering
    \includegraphics[width=0.5\linewidth]{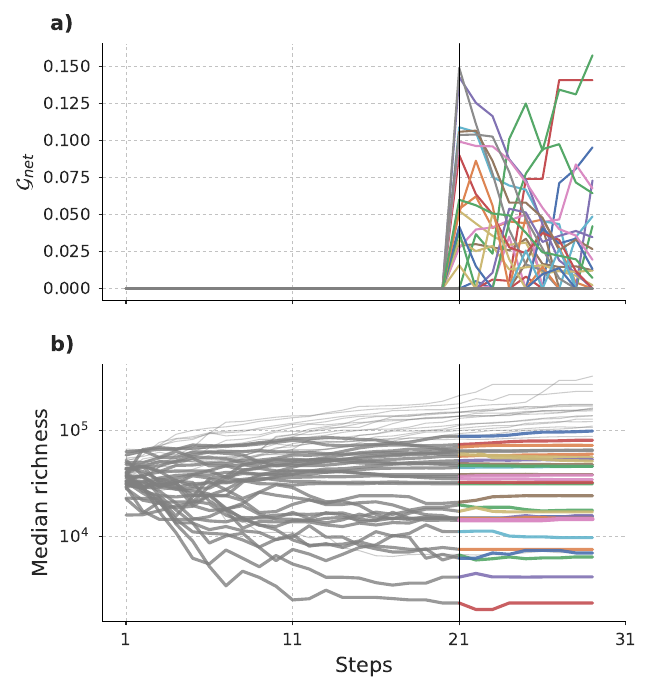}
    \caption{\textbf{a)} Time series of $\mathcal{G}_{net}(t,\Delta)$, with fixed values of the time window width $\Delta=15$, of only the nodes that undergo sharp peaks of $\mathcal{G}_{net}$. \textbf{b)} Time series of the richness (median of agents' income) of all nodes in the network: the colored curves correspond to the same nodes whose time series of $\mathcal{G}_{net}$ is presented in \textbf{a)}. The black vertical line in both plots indicates the initial time step when $H$ agents become eligible to relocate, occurring $\theta$ steps after the simulation's start.}
    \label{fig:random_richness_gnet}
\end{figure}

\begin{figure}[H]
    \centering
    \includegraphics[width=0.9\linewidth]{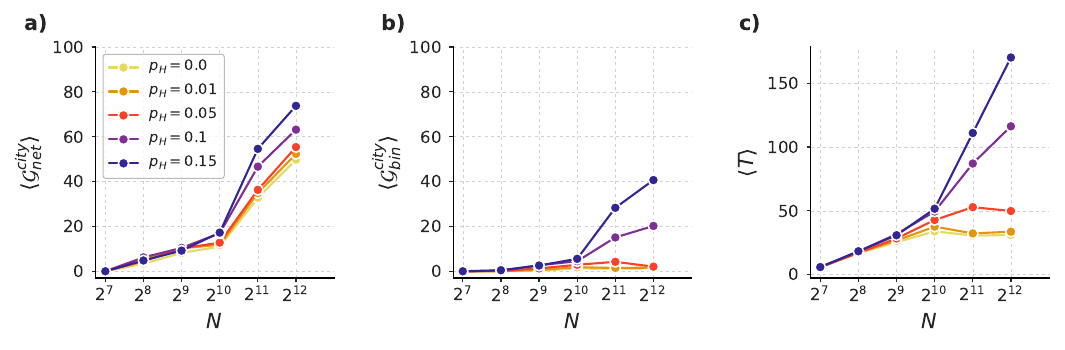}
    \caption{Average results across 150 simulations for varying agent populations (x-axis, logarithmic scale) and high-income agent relocation rates $p_H$ (colors). \textbf{a,b)} City-wise gentrification levels measured by $\mathcal{G}_{net}^{city}$ (a) and $\mathcal{G}_{bin}^{city}$ (b). \textbf{c)} Number of average simulation time steps. Higher urban densities facilitate gentrification and increase convergences time.}
    \label{fig:city_levels_numbers}
\end{figure}

The striking difference between the core-periphery initial configuration of the city, and the random initial spatial distribution of agents lies in the whole-city gentrification level analysis: in fact, while for low values of $p_H$ $\mathcal{G}_{bin}$ does not capture any gentrification events (Figure \ref{fig:random_boxplots}.a), $\mathcal{G}_{net}$ does, even for $p_H=0$ (Figure \ref{fig:random_boxplots}.b). As we see in Figure \ref{fig:random_richness_gnet} this is due to the fact that even if H agents are not moving, both M and L agents are immediately moving, so, only after the buffer $h=20$, $\mathcal{G}_{net}$ captures as true gentrification events the neighbourhood transitions that unfold in the very few steps of the simulation. Interestingly we see how with the random initial distribution of agents, the model seems to rapidly converge to a segregated, final configuration, more so than the core-periphery structure. This is evident in Figure \ref{fig:city_levels_numbers}, in particular in Figure \ref{fig:city_levels_numbers}.c, where we note that the average time required to reach the termination condition is consistently lower than what shown in the main text for all values of the number of agents in the city.

\section{Neighbourhood impoverishment}
We introduce a measure capable of capturing the outflow of M and H agents as well as the simultaneous inflow of L agents within a neighbourhood. This is achieved analogously to what is done for $\mathcal{G}_{net}^i(t,\Delta)$ in equations 11-13, but now considering the net-outflow of H and M agents $\varphi^{out}_{M+H,i}(t)$ and the net-inflow of L agents $\varphi^{in}_{L,i}(t)$:

\begin{equation}
\mathcal{D}_{net}^i(t,\Delta)\equiv \sqrt{\Bigg(\frac{1}{\Delta} \sum_{\tau=t-\Delta}^{t} \varphi^{\text{out}}_{M+H,i}(\tau)\Bigg) \cdot \Bigg(\frac{1}{\Delta} \sum_{\tau=t-\Delta}^{t} \varphi^{\text{in}}_{L,i}(\tau)\Bigg)}.
    \label{eq:Dnet}
\end{equation}

\begin{figure}[H]
    \centering
    \includegraphics[width=0.9\linewidth]{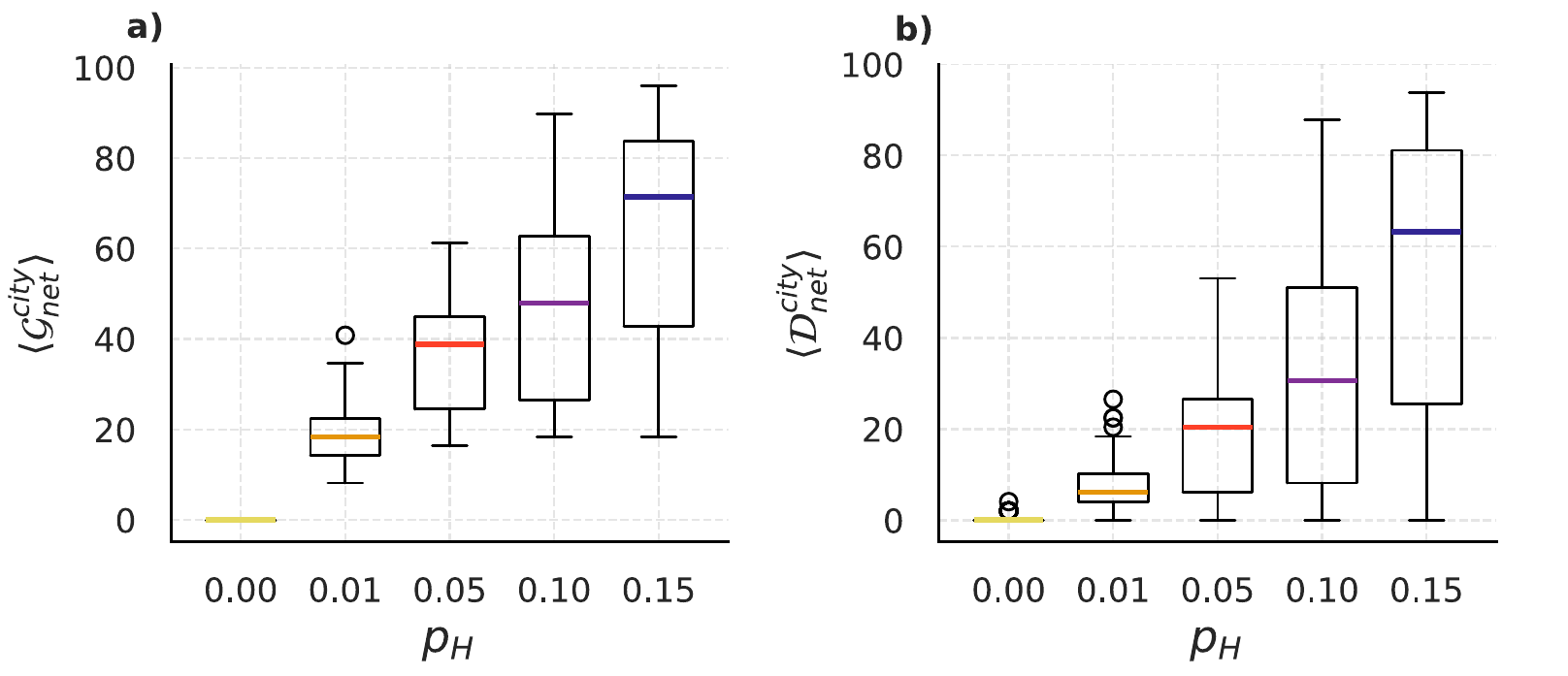}
    \caption{\textbf{Gentrification vs. Impoverishment:} \textbf{a)} Gentrification level across 150 simulation of the model with a fixed $\theta=20$, $N=2^{12}$ and varying value of $p_H$ according to and to the network based measure $\mathcal{G}_{net}^{city}$. \textbf{b)} Impoverishment level across 150 simulation of the model with a fixed $\theta=20$, $N=2^{12}$ and varying value of $p_H$ according to and to the network based measure $\mathcal{D}_{net}^{city}$.}
    \label{fig:DvsG_boxplots}
\end{figure}

\begin{figure}[H]
    \centering
    \includegraphics[width=0.5\linewidth]{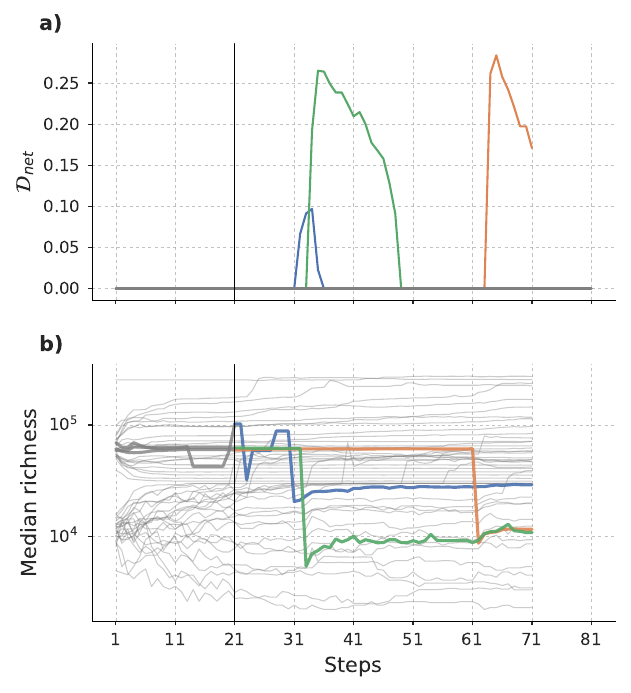}
    \caption{\textbf{$\mathbf{\mathcal{D}_{net}}$ peaks capture richness drops in neighbourhoods:} \textbf{a)} Time series of $\mathcal{D}_{net}(t,\Delta)$, with fixed values of the time window width $\Delta=15$, of only the nodes that undergo sharp peaks of $\mathcal{D}_{net}$. \textbf{b)} Time series of the richness (median of agents' income) of all nodes in the network: the colored curves correspond to the same nodes whose time series of $\mathcal{D}_{net}$ is presented in \textbf{a)}. The black vertical line in both plots indicates the initial time step when $H$ agents become eligible to relocate, occurring $\theta$ steps after the simulation's start.}
    \label{fig:D_richness}
\end{figure}

This measure of impoverishment, or disinvestment of a neighbourhood, thus captures \emph{downward} transitions of neighbourhoods in terms of income distributions. Similarly to the analyses presented in Figure 3 in the main text, we compute the fraction of neighbourhoods displaying peaks of $\mathcal{D}_{net}$ with different values of the moving probability of H agents $p_H$. We show a direct comparison of the fraction of neighbourhoods undergoing peaks of $\mathcal{G}_{net}$ with that of neighbourhoods undergoing peaks of $\mathcal{D}_{net}$ in Figure \ref{fig:DvsG_boxplots}. Here we note that the number of $\mathcal{D}_{net}$, i.e., the overall number of impoverishment events across the city, increases with $p_H$, even though with lower numbers w.r.t $\mathcal{G}_{net}$ and with a much higher variance, hinting that even with higher $p_H$, impoverishment events do not happen in some simulations. We display an exemplary plot of the time series of $\mathcal{D}_{net}$ (obtained in one simulation) in Figure \ref{fig:D_richness}.a. Here we note how the peaks of $\mathcal{D}_{net}$ occur in accordance with radical, downward shifts of the richness in the corresponding neighbourhoods (Figure \ref{fig:D_richness}.b).

\section{Analysis of longer simulations (termination condition removed)}

\begin{figure}[H]
    \centering
    \includegraphics[width=0.5\linewidth]{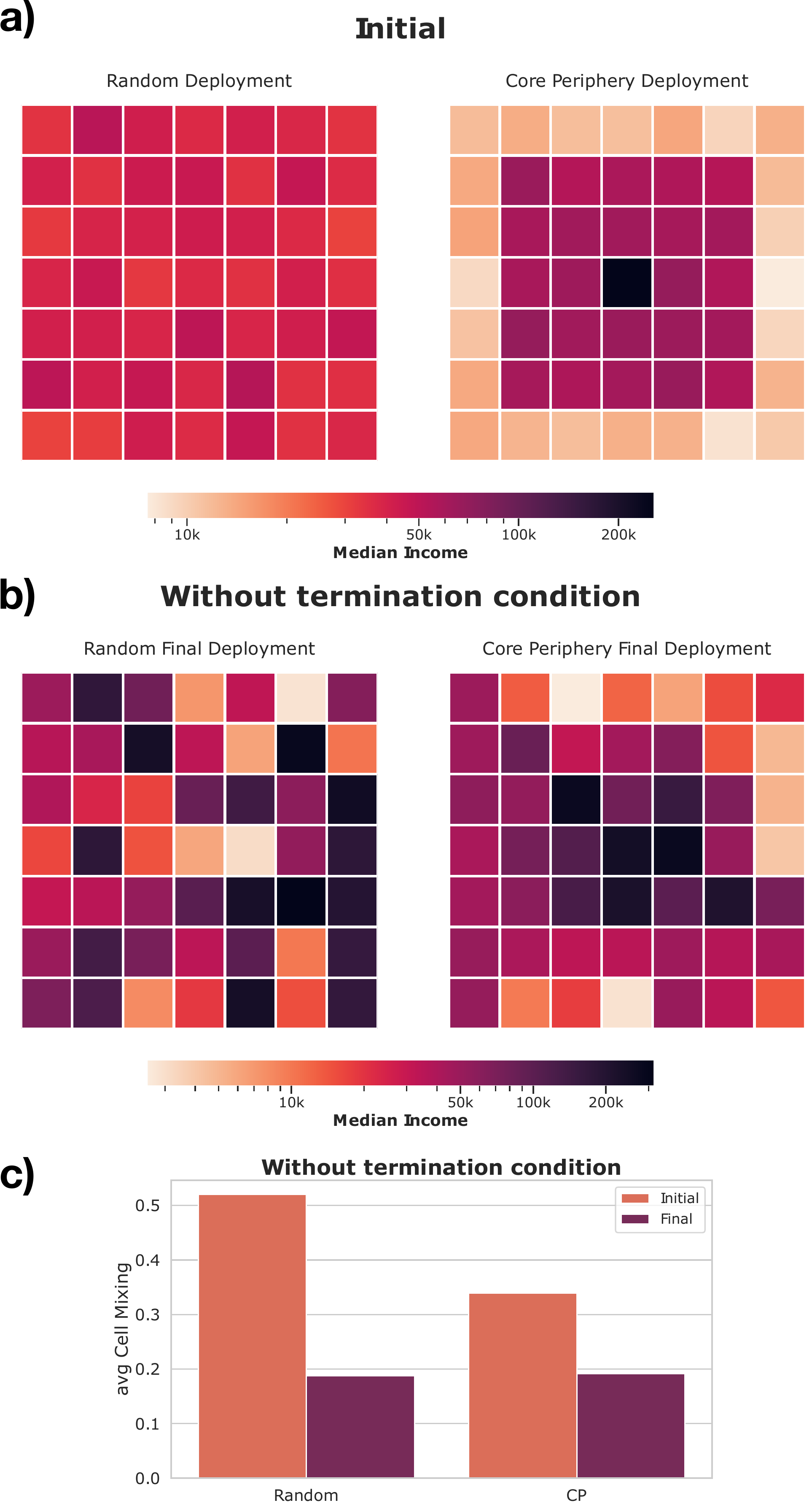}
    \caption{\textbf{Random vs Core-Periphery initial configuration without termination condition:} \textbf{a)} heatmaps of neighbourhood median income at the beginning of the simulation with a random placement of agents, i.e., each neighbourhood has a fraction of agents of the three types close to the fraction of agents of each type in the whole city-grid (left) and of the mono-centric, core-periphery configuration of the city as discussed in the main text (right). \textbf{b)} same heatmaps but obtained from the last step of the simulation of the model, \textbf{without} the termination condition: thus the model runs until a prefixed limit of 2000 steps. \textbf{c)} On the y-axis is a measure of the average mixing of agents in each neighbourhood, i.e., the Gini index computed over the distribution of income within each grid-cell, averaged over all cells. The orange bars correspond to the value computed for the initial configuration (both for random or core-periphery spatial distribution of agents), while the purple bar corresponds to the same value computed at the last step of the corresponding simulation.}
    \label{fig:mixing_long}
\end{figure}

\begin{figure}[H]
    \centering
    \includegraphics[width=\linewidth]{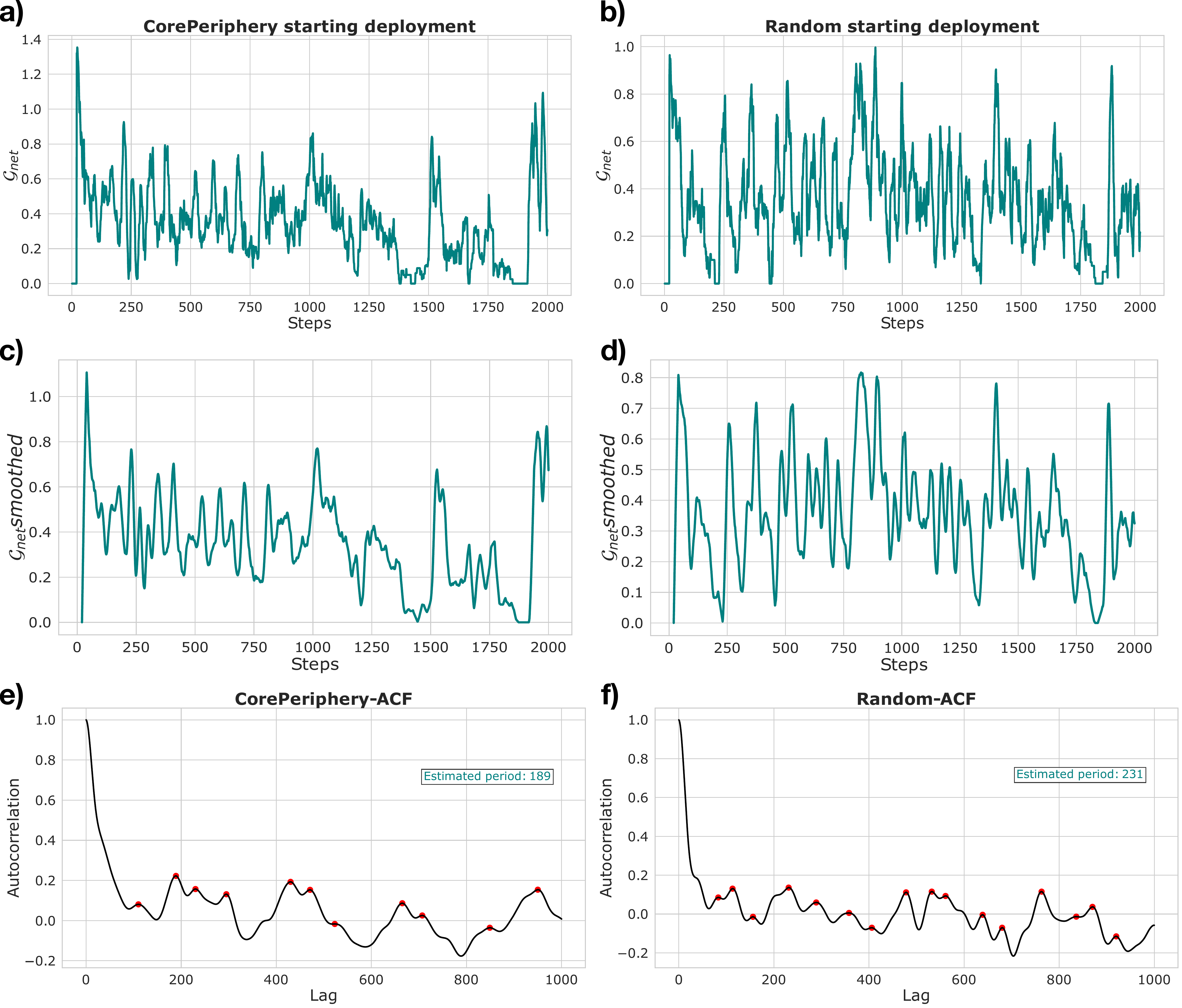}
    \caption{\textbf{Periodicity of gentrification events:} \textbf{a,b} Time series of global $\mathcal{G}_{net}$ (sum over all neighbourhoods) in longer simulations (2000 steps) with core-periphery (a) and random (b) initial configuration. \textbf{c,d} Smoothed time series of global $\mathcal{G}_{net}$ (sum over all neighbourhoods) with a rolling window of width $20$ steps in longer simulations (2000 steps) with core-periphery (c) and random (d) initial configuration. \textbf{e,d} Autocorrelation function at lags $\text{Lag}\in[0,1000]$ for the smoothed time series of global $\mathcal{G}_{net}$ (sum over all neighbourhoods) measured for the core-periphery initial configuration (e) and for the random initial configuration (f), estimated period, i.e., highest peak of the autocorrelation function, in the inset.}
    \label{fig:ACF}
\end{figure}

When allowing for longer simulation, i.e., removing the termination condition and, instead, setting a limit of 2000 steps to the model, we find that the final configuration (at $t=2000$) obtained by both the random and core-periphery initial deployments of agents, are more segregated (lower mixing level) than what is found with the termination condition (Figure \ref{fig:mixing_long}). In order to investigate the periodicity of gentrification events or waves in both versions of the model, without the termination condition, we compute the auto correlation function (ACF) of the time series of global $\mathcal{G}_{net}$, i.e., the sum over all neighbourhoods $\sum_{i\in[0,N]}\mathcal{G}_{net}^i$ (Figure \ref{fig:ACF}.a,b). To avoid biasing of this analyses due to very fast fluctuations of the time series, we compute the ACF for a smoothed version of the time series (Figure \ref{fig:ACF}.c,d), where we employ a time window of width $20$ (same as $\Delta$ used for the computation of $\mathcal{G}_{net}$ (Figure \ref{fig:ACF}.e,f). We note how the highest peaks of the ACF for both initial agent deployment strategies lie close to 200 steps (189 for core-periphery, 231 for random). 

